\documentclass[structabstract]{aa}  
\usepackage{graphicx}
\usepackage{rotating}
\usepackage{pdflscape}
\usepackage{subfigure}
\usepackage{txfonts}

\usepackage{color}

\usepackage{natbib}
\usepackage{enumerate}
\bibpunct{(}{)}{;}{a}{}{,} 

\begin{document}

\title{Metallicity gradients in Local Universe galaxies: \\
time evolution and effects of radial migration}

\author{Laura Magrini\inst{1}, Lodovico Coccato\inst{2}, 
Letizia Stanghellini\inst{3}, Viviana Casasola\inst{1}, Daniele Galli\inst{1}}


\institute{
1-INAF -Osservatorio Astrofisico di Arcetri, Largo E. Fermi, 5, I-50125 Firenze, Italy\\
2-ESO Karl-Schwarzchild str., 2,  85748 Garching b. Munchen, Germany\\
3-National Optical Astronomy Observatory, 950 N. Cherry Avenue, Tucson, AZ 85719, USA\\
\email{laura@arcetri.astro.it} }

\date{Received ; accepted }

   \abstract
   {Our knowledge of the shape of radial metallicity gradients in disc galaxies has recently  improved. Conversely, the understanding of their time evolution is more complex, since it requires analysis of stellar populations with different ages, or systematic studies of galaxies at different redshifts.  In the Local Universe, H{\sc ii} regions and planetary nebulae (PNe) are important tools to investigate it.  }
{We present an in-depth study of all nearby spiral galaxies (M33, M31, NGC300, and M81) with  {\em direct}-method nebular abundances of both populations, aiming at studying the evolution of their radial metallicity gradients. For the first time, we also evaluate the radial migration of PN populations.  }
{For the selected galaxies, we analyse H{\sc ii} region and PN properties to: determine whether oxygen in PNe is a reliable  tracer for past interstellar medium (ISM) composition;
homogenise the published datasets; estimate the migration of the oldest stellar populations; determine the overall chemical enrichment and slope evolution.}
{We confirm that oxygen in PNe is a reliable tracer for the past ISM metallicity.  We find that PN gradients are flatter than or equal to those of H{\sc ii} regions. 
When radial motions are negligible, this result provides a direct measurement of the time evolution of the gradient. For galaxies with dominant radial motions, we provide upper limits on the gradient evolution. Finally, the total metal content increases with time in all target galaxies, with early morphological type having a larger increment $\Delta$(O/H) than late-type galaxies. }
{Our findings provide important constraints to discriminate among different galactic evolutionary scenarios, favouring cosmological models 
with enhanced feedback from supernovae. 
The advent of extremely large telescopes will allow us to include galaxies in a wider range of morphologies and environments, thus putting firmer constraints to galaxy formation and evolution scenarios.}

\authorrunning{Magrini, L. et al.}
\titlerunning{\sc Time evolution of radial metallicity gradients}

\maketitle

\section{Introduction}

Spiral galaxies are complex astrophysical objects, showing a
non-uniform distribution of metals across their discs.  Their radial
metallicity distribution, known as the radial metallicity gradient,
has been studied for a long time, starting with the 
pioneering works  by  \citet{aller42} and later by \citet{searle71}
and \citet{pagel81}.  Nowadays the existence of radial 
metallicity gradients in spiral galaxies has been tackled 
in the following cases: individual nearby galaxies,  for which {\em
direct}-method measurements are available, i.e., abundance analysis that involves the
measurement of the electron temperature and density diagnostic emission
lines to characterise the physical conditions of the plasma \citep[see, e.g.,][]{bresolin07,bresolin09,berg12,
berg13,berg15}; large samples
of  intermediate distance galaxies,  where H{\sc ii} region
abundances are derived  from the detection of several ``strong''
emission lines \citep[see, e.g.,][]{sanchez14}; and high-redshift
galaxies \citep[see, e.g.,][]{cresci10, jones10, jones13, jones15}.

In our Galaxy, metallicity gradients  have been estimated using
several tracers, from those of relatively young ages, such as OB
stars \citep[e.g.,][]{daflon04}, Cepheids \citep[e.g.,][]{Andrievsky04,
luck06,  yong06}, H{\sc ii} regions \citep[e.g.,][]{Deharveng00, esteban05,
rudolph06, balser11}, to those of intermediate-to-old ages, such as
planetary nebulae \citep[e.g.,][]{maciel03, henry10, stanghellini10}
and open clusters \citep[e.g.,][]{friel95, cheng03, magrini09}.

The general result is that most spiral galaxies show negative radial
gradients within their optical radius.  In addition, when the
observations are deep enough to investigate the outskirts of galaxies,
it has been often found that the metallicity tends to reach a plateau: in external galaxies
whose gradients are marked by H{\sc ii} regions
\citep{bresolin09, goddard11,  werk11, bresolin12}   and in our own Galaxy
using open clusters \citep{sestito06,magrini09,lepine11},  
H~{\sc ii} regions \citep{VE96}, and PNe \citep{MQ99}. 
\footnote{It is worth noting that not all Galactic tracers indicate that the radial metallicity gradients are flat in the outer parts of the galactic disc. 
For example, Cepheid gradients \citep{luck11, lemasle13} seem to be negative at all radii.
 }

If we focus  on  regions within the galactic
optical radius, where the gradients are clearly negative, the most
important open questions are: how was the present time radial
gradient  established? Was it steeper or flatter in the past? Finding answers
to these questions has important implications to our understanding
of the processes that lead to the disc formation.  
In particular, observations allow us to  put strong constraints on
galactic formation and evolution models  \citep[e.g.,][]{pilkington12a,pilkington12b, gibson13,
stinson13}.

In classical chemical evolution models -- in which the cosmological
context and the gas and star dynamics are neglected -- the different
predictions on the time evolution of the radial metallicity gradients
depend on the pre-enrichment of the material that  formed the
primordial disc and whether one assumes a gas density threshold for star formation. 
A disc formed
by pre-enriched gas, and in which  a minimum gas density is required
to permit the formation of new stars, naturally develops an initial
flat metallicity gradient that becomes steeper with time \citep[see,
e.g.,][]{chiappini01}.  On the other hand, in models in which the
disc is formed by primordial gas and the star formation can proceed
at any gas density, the radial metallicity gradient is typically
steeper at early times, and flattens as the galaxy evolves
\citep{ferrini94,hou00, molla05, magrini09}.
Recently, \citet{gibson13} have examined the
role of energy feedback in shaping the time evolution of abundance
gradients  within a sub-set of cosmological hydrodynamical disc
simulations drawn from the MUGS \citep[McMaster Unbiased Galaxy
Simulations;][]{stinson10} and MaGICC \citep[Making Galaxies in a
Cosmological Context;][]{brook12} samples.  The two sets of
simulations adopt two feedback schemes:  the
conventional one in which  about 10-40\% of the energy associated
with each supernova (SN) is used to heat the ISM,
 and the enhanced feedback model in which a larger quantity of
energy per SN is released, distributed and re-cycled over
large scales. The resulting time evolution of the
radial gradients is different in the two cases: a strong flattening with time in the former, and 
relatively flat and temporally invariant abundance gradients in the latter.

In addition, the modelling of the chemical evolution of the galactic disc needs
to be combined with dynamics.  Both observations and numerical
simulations show that stars move from their birth places and
migrate throughout the disc during their lifetimes. In our Galaxy, the age-metallicity 
relationship in the solar neighbourhood puts
strong constraints on the presence of radial migration
\citep{edvardsoon93, haywood08, schonrich09, minchev10}: 
for any given age, there are stars in the solar neighbourhood with a wide range of metallicities 
and this can only explained with an exceptionally inhomogeneous ISM or, more likely, with radial migration. 
From a theoretical point of view, radial stellar migration is related
to several processes as the presence of transient spirals
\citep{sellwood02, roskar08}, of  mergers and perturbations from satellite galaxies and sub-haloes that can induce radial mixing  
\citep{quillen09}, and  of a central bar that  can 	stimulate
a strong exchange of angular momentum when associated to a spiral
structure \citep{minchev10,  minchev11,  dimatteo13, Kubryk+13}.  \citet{minchev14},
by separating the effects of kinematic heating and radial migration,
showed the importance of   migration needed,  especially,  to explain the oldest and hottest stellar population.

From an observational point of view, the  time evolution of the radial metallicity
gradient in nearby external galaxies  is well studied with two tracers
with similar spectroscopic features, characterising to two different
epochs in a galaxy lifetime: H{\sc ii} regions and PNe. Studies of
PNe and H{\sc ii} region gradients from {\em direct}-method
abundances in M33 \citep{magrini09b, magrini10, bresolin10},
M81 \citep{S10, S14}, NGC300 \citep{bresolin09b, stasinska13}, and M31 \citep{ZB12}  
have produced a limited array of results, all indicating
gradient invariance, or steepening of the radial O/H gradient
with time.  These results, summarised in Fig.~9 of \citet{S14}, need to
be validated by a thorough and homogeneous analysis of all the
available data and  the gradients recalculated over a common
galactocentric scale for all sample galaxies.

The paper is organised as follows: in Section~\ref{sec1} we prove the reliability of PN abundances to investigate the past ISM composition through the invariance
of O (at first order) during the life-time of PN progenitors, while in Section~\ref{sec2} we present our sample of spiral galaxies with the observational datasets. 
In Section~\ref{sec:migrating} we 
estimate the effects of radial migration in M31 and M33. In Section~\ref{sec4} we describe  the adopted
method to homogenise the literature data and to study the time-evolution of radial metallicity gradients.  
 In Section~\ref{sec5} we summarise our results, and in Section~\ref{sec6}  we discuss them. 
 We give our summary and conclusions in Section~\ref{sec7}.

\section{Tracing the past: the ISM composition}
\label{sec1}

H{\sc ii} regions are among the best tracers of the present-time composition
of the ISM since they are ionised by young massive
OB stars that  did not have enough time to move from their birth
place.  On the other hand, PNe are the gaseous remnants of
relatively old stellar populations. Before using them as tracers
of the past ISM composition, we must consider the nucleosynthesis
and mixing processes taking place during their evolution that could modify
their composition, and the relevance of their dynamics that could 
displace them from their initial place of birth (see Section~\ref{sec:migrating}).

To obtain a reliable determination of the 
chemical composition of a nebula from emission lines, we need accurate measurements of the 
nebula's physical
properties, such as electron temperature $T_e$ and density
$N_e$. Elemental abundances are often derived from the measurement
of collisionally excited lines (CELs), which are very sensitive to
$T_e$. The latter is derived from the CEL ratios,
such as, for instance, [O{\sc iii}] $\lambda$4363/($\lambda$4959 +
$\lambda$5007),  and [N{\sc ii}] $\lambda$ 5755/($\lambda$6548 + $\lambda$6584)
nebular-to-auroral line ratios \citep{OF06}.  

An alternative method for nebular abundance determinations is to use
the  ratio of the intensity of an optical recombination line (ORL) of
He or a heavy element, such as O or N,  with that of H.  The ORLs are less
affected by temperature measurement errors since they have a weak
dependence on $T_e$ and $N_e$.  In regions where both CEL and ORL
abundances are available, the ORLs give systematically higher
abundances than CELs \citep[e.g.,][]{peimbert93,  liu95, liu01,
tsamis04,ge07}. This behaviour has been explained, for instance,  with the bi-abundance
nebular model by \citet{liu00} in which the ORLs abundances arise
from cold H-deficient small portions of the nebulae, while the
strong CELs are emitted predominantly from the warmer ionised gas,
and thus are more representative of the global nebular abundance.  In
addition, ORLs are much weaker than CELs, and consequently they can
be used to measure abundances only in nearby or bright sources.
Thus, in most extragalactic studies CELs are adopted to trace
the chemical composition,  and the total abundance of a given element is computed by
summing the abundances of the different ionisation stages of that element.   

However, not all the ions are observable in the optical range, 
and since almost all abundance
measurements are based on optical spectroscopy, the ionic sum
may underestimate the atomic abundances.  The contribution of the
unseen ions is usually estimated introducing the so-called ionisation correction factors
(ICFs), often based  on photoionisation models \citep{KB94, KH01,
delgado14}.  In this framework, O is the best measured
element in PNe (and H{\sc ii} regions) because: ({\em i}) we can directly measure
its electron temperatures $T_e$([O{\sc iii}])  from  [O{\sc iii}]
$\lambda$4363/($\lambda$4959 + $\lambda$5007)  and $T_e$([O{\sc ii}])
from  [O{\sc ii}] $\lambda$7325/($\lambda$3727); ({\em ii}) the 
transitions relative to the most abundant ionisation stages, [O{\sc i}],
[O{\sc ii}], and [O{\sc iii}],  are available in the optical range, and
consequently no correction for the unseen ionisation stages is
needed for the most common low- and intermediate-excitation
PNe. In PNe of higher ionisation, the contribution of other O
ions may be significant \citep[see][for an updated treatment of the
ICF schema]{delgado14}.

Before using O/H abundance in PNe as a tracer of the past ISM composition in
spiral discs, one needs to prove, from an observational point of view,
that O has not been modified during the PN progenitor
lifetime in the metallicity range of our interest, typically [12+$\log ({\rm O/H})$]$>8.0$.
The production of O and Ne is indeed dominated by Type II
supernovae whose progenitors are massive stars with $M>8$~$M_\odot$
\citep{WW95, herwig04, CL04}.  From stellar evolution of low- and
intermediate-mass stars, we know that the abundance of $^{16}$O can
be slightly reduced as a consequence of hot bottom burning in the
most massive progenitors.  On the other hand, low-mass stars may
have a small positive yield of $^{16}$O \citep{marigo01, karakas07}
at low metallicity, while the same effect is negligible for the
same stars at solar metallicity.  Thus, the O abundance is,
in general,  expected to be little affected by nucleosynthesis in
PN progenitors.
The same is true for the abundances of other $\alpha$-elements, such as Ne,
Ar, and S. 
A complete review of the sites and processes of production of Ne
and O in PNe is presented in \citet{RM08}.  

\begin{figure}
\centering
\includegraphics[width=0.35\textwidth, angle=270]{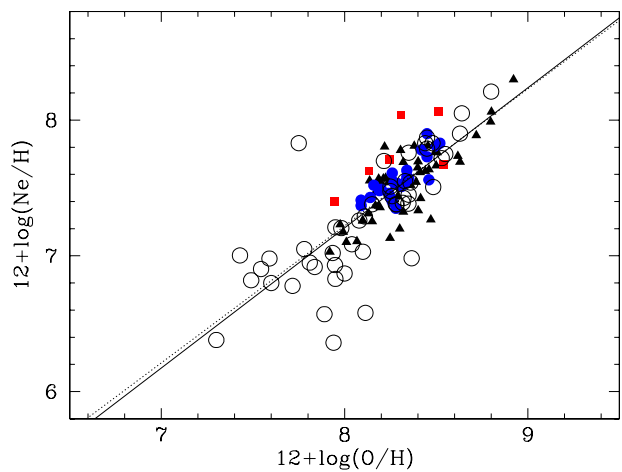}
\caption{Plot of $12+\log({\rm O/H})$ vs. $12 + \log({\rm Ne/H})$ in PNe in
M33 (black triangles), M81 (red squares), and NGC300
(blue filled circles).  PNe in Local Group dwarf galaxies are shown
as empty circles. The continuous line is the linear fit of the
abundances of PNe in the whole sample of galaxies, while the dotted
line is the fit by \citet{izotov06} of star-forming galaxies.}
\label{OHNeH}
\end{figure}

From the observational point of view, a study of the relation between Ne and O for a large range
of metallicities provides a good way to prove their invariance during the lifetime of PN progenitors.
In Fig.~\ref{OHNeH} we plot O/H vs. Ne/H
in a sample of PNe belonging to a variety of galaxies in the Local
Group: IC 10 \citep{MG09}, Sextans~A and
Sextans~B \citep{magrini05}, Leo A \citep{vanzee06}, NGC3109
\citep{pena07}, NGC6822 \citep{hernandezmartinez09}, NGC205
\citep{RM08, goncalves14}, NGC185 \citep{goncalves12}, and NGC147
\citep{goncalves07}. The sample also includes 
three of the spiral galaxies studied in the present paper:  M33 
\citep{magrini09b, bresolin10}, NGC300 \citep{stasinska13}, 
and M81 \citep{S10}. \footnote{For M31 only the O
abundance is available in \citet{sanders12}}. 
From Fig.~\ref{OHNeH}, we
can see that O and Ne abundances are strictly coupled in
the majority of PNe, with a slope 
close to the unity (1.03$\pm$0.07).  Comparing with the same relationship obtained for
H{\sc ii} regions  \citep[with slope 1.01,][]{izotov06}, we confirm that the two relations
are very similar, with  insignificant difference in their slopes, probing that O and Ne in PNe can be used to safely trace 
the past composition of the ISM.

\subsection{Dating the PN populations} 
When using PNe as indicators of the chemical evolution of galaxies,
one should be aware that different PN populations probe different
epochs, and are subject to different selection effects.  
Historically, Galactic PNe have been divided into three classes,
based on their N and He content, on their location with
respect to the Galactic plane, and on their radial velocity \citep[for a 
summary and references, see][]{stanghellini10}. Type I PNe
are highly enriched in N and He, an occurrence for only
the asymptotic giant branch (AGB) stars with an initial mass larger than 2--4.5~$M_\odot$
depending on metallicity \citep[e.g.,][]{karakas10}, corresponding to a
very young progenitor population. All non-Type I PNe with high
radial velocities are defined as Type III: they are typically located
in the Galactic halo, and they are a small minority of Galactic
PNe, representing the progeny of the lowest-mass AGB stars. Finally,
the most common PNe in the Galaxy are Type II PNe, with intermediate
AGB mass progenitors.

However, it is not possible to determine the exact range of progenitor mass 
for the PNe of each class. From observations, the minimum
progenitor mass for a Type I PN is $\sim2$~$M_\odot$ \citep{PS80}, 
and for Type II PN is 1.2~$M_\odot$ \citep{perinotto04}. 
From stellar evolution models, the mass range seems to be
correlated to metallicity, with the Type~I PN mass cutoff (i.e.,
the critical mass for hot bottom burning) decreasing
with metallicity. The population of extragalactic PNe is very
similar to that of Galactic PNe, apart from a weak metallicity
shift in the limit mass for Type~I PN progenitors.
The initial mass function (IMF) favours lower mass stars, thus  Type~I PNe are 
generally scarce in any galaxy. Furthermore, the evolution of the most massive 
post-AGB stars is very fast, making the corresponding PN shining for only
a few years. Therefore, Type~I PNe in external galaxies are a very small
young population contaminant to the whole PN sample.
Type~III PNe are typically not observed in spiral galaxies, 
since they do not trace the spiral arms, where
spectroscopic targets are typically chosen. Furthermore, in
extragalactic samples one normally observes only the brightest first or second bins
of the PN luminosity function (PNLF), which excludes 
Type~III PNe (because they do not reach high luminosity), so most
samples are populated by Type~II PNe.
In the present work we select non-Type I PNe from the original
samples, thus most PNe have progenitor mass in the range $1.2<M<2$~$M_{\odot}$.
This places our PNe at $\sim 1$--5~Gyr look-back time \citep{maraston98}.

\section{The sample}
\label{sec2} 
There are only four spiral galaxies (other than the Milky Way)
in which both H{\sc ii} region and PN abundances are obtained with the
{\em direct}-method.  They constitute our sample.  The main properties of
the four galaxies, including coordinates, distance, optical radius,
inclination, position angle, heliocentric systemic velocity and morphology,  are listed in Table~\ref{tab_gal}.  
The references for distance and optical radius are quoted in the Table, while for inclination, position angle,  morphology we adopt the values from HyperLeda\footnote{http://leda.univ$-$lyon1.fr/}.  
For some galaxies, the heliocentric systemic velocity is determined as described in Section~\ref{sec:migrating1}. The quoted value is the average of the velocities obtained from all the individual ellipses. For other galaxies, we the value is obtained from HyperLeda.
The
sample galaxies have morphological types from very late
(NGC300 - Scd) to early type (M81- Sab).  M31 and M33 belong to the
Local Group: M33  does not have strong interactions with dwarf 
companions, although it might  have had some mutual interaction 
with M31 in the past \citep{mcconnachie09, putman09}. 
M31 has several companions that might perturb its dynamics \citep{chemin09, dierickx14}.
NGC300 is quite isolated, though traditionally considered as a part of the 
Sculptor group.
Finally, M81 is the largest  galaxy of the
M81 group and it is strongly interacting with its two brightest companions,
M82 and NGC3077, located within a short projected distance  \citep[60~kpc, ][]{kaufman89}. The distribution of atomic hydrogen shows several extended
tidal streams between M81 and its companions \citep{yun94,allen97}, which
might be the result of close encounters $\sim 200$--300 Myr ago.
In addition, M81 presents other peculiarities, as a reduced  content of molecular gas with respect to other 
spiral galaxies \citep[][]{casasola04,casasola07}.

The PN and H{\sc ii} region populations have been previously studied
via medium-resolution optical spectroscopy in all galaxies of our
sample.  In the following we select abundance determinations via {\em direct} method for the reasons discussed in Section~\ref{sec1}. 
In order to produce a homogeneous sample of abundances, we selected  spectroscopic observations all obtained with comparable telescopes and spectral resolutions. 
The details of the observations are shown in 
Table~\ref{tab_obs}, including the reference for the determination of abundances, the telescope name (William Herschel Telescope --WHT; Multiple Mirror Telescope --MMT; Very Large Telescope --VLT) with its diameter, the spectral resolution, the signal-to-noise ratio (SNR) of the 
faintest auroral lines, and the typical uncertainties on the O abundance. In all datasets, the SNR is adequate for an accurate abundance analysis. 
In each galaxy, the uncertainties on O abundances from different literature sources are all comparable.   

For M33 we adopt abundances of PNe from \citet{magrini09b} (M09)
and from \citet{bresolin10} (B10), while for H{\sc ii} regions from \citet{magrini07,
magrini10} (M07, M10) and references therein, and B10.  
For
M31, the largest and most recent sample of H{\sc ii} region abundances
is computed by using strong-line ratios \citep{sanders12} (S12). For this galaxy
we prefer to take advantage of the large sample of \citet{sanders12} of  H{\sc ii} region abundances
and to use the sample \citep{ZB12} (ZB12) with {\em direct}-method  abundances to
compute the offset between the {\em direct} method and the N2=[N{\sc ii}]/H$\alpha$
calibration.  
For PNe, there are several studies focussed on different PN populations of M31: 
PNe in the bulge and disc of M31 \citep{JC99}, in the very outer disc \citep{kwitter12, balick13, corradi15}, 
in the Northern Spur and the extension of the Giant Stream \citep{fang13, fang15}, and in the whole disc  \citep{sanders12}.
Since we are interested in the behaviour of the gradient within the optical radius, we consider here the abundance determination 
of \citet{sanders12},  based on {\em direct} method, including only PNe that kinematically belong  to the disc population, i.e., excluding halo and satellite objects. 
There are several studies of  H{\sc ii} region abundances in M81, some based 
on strong-line \citep{GS87, bresolin99} and  others based on the {\em direct} method  \citep{S10, patterson12, S14, AC15}. 
We adopt the {\em direct}-method results from \citet{S10,S14} (S10, S14), and from \citet{patterson12} (P12).
In the case of \citet{S14}'s sample, following their analysis, we included only H{\sc ii} regions with uncertainties in O
abundance $<0.3$~dex in our sample.
\footnote{ 
We excluded from our sample the  abundances of \citet{AC15} because of an alternate approach in deriving oxygen abundances, with different temperatures for different ions 
based on a relation between [O{\sc iii}] and [N{\sc ii}] T$_{e}$ \citep[e.g.,][]{esteban09}, with respect to \citet{S10,patterson12,S14} who used only measured temperatures for all ions.
Both approaches are, in principle, correct, but they can produce divergent  results when  the observational limits  of the auroral lines are reached, as in M81.}
The only {\em direct}-method abundance determinations available for M81 PNe are those from \citet{S10}, which we use in this paper.
Finally for NGC300, the O/H abundances
of PNe and of H{\sc ii} regions are  from \citet{stasinska13} (S13) and
\citet{bresolin09b} (B09).  

For all PN samples, since we are interested in the time
evolution of the radial metallicity gradient, we have excluded, 
whenever possible, Type I PNe,
keeping the PNe with the older progenitors.

\begin{table*}
\caption{Properties of the galaxy sample}
\tiny
\begin{tabular}{lllllllll}
\hline
Galaxy &  RA & Dec & Distance & $R_{25}$ & Inclination   		& PA & SysVel  & Morphology\\
 	   &\multicolumn{2}{c}{(J2000.0)} & (Mpc) & (kpc) & (deg)   & (deg)& (km~s$^{-1}$) &		      \\
\hline
NGC300 	& 00h54m53.48s & +37d41m03.8s  &  1.88 \citep{bono10}	         & 5.3 \citep{bresolin09b} 	& 48.5 	& 114.3 & +165 & Scd 	\\
M33  	& 01h33m50.9s   & +30d39m37s     &  0.84 \citep{freedman01}	         & 9.0 \citep{magrini07}		& 55.0    	& 22.7  &  --185$^{a}$ 	& Sc  	\\
M31 	        & 00h42m44.3s   & +41d16m09s     &  0.77  \citep{tully13}	                 & 20.6 \citep{draine14}    	& 72.2 	& 35.0  & --325$^{a}$   &Sb 	\\
M81 	        & 09h55M33.2s   & +69d03m55s     &  3.63  \citep{gerke11} 	         & 14.6 \citep{patterson12} 	& 62.7	& 157.0 & -38	 &Sab 	\\
\hline
\end{tabular}
\label{tab_gal}
\begin{minipage}{21cm}
Notes -- Data with no explicit references are taken from HyperLeda. $^{a}$ Determined from the fit in Section \ref{sec:migrating1}.
\end{minipage}
\end{table*}

 \begin{table*}
\begin{center}
\caption{Details of the observations of the considered samples.}
\tiny
\begin{tabular}{lllllll}
\hline
Galaxy &  Ref. & Telescope & Diameter & Resolution & SNR  & $\delta$(O/H)\\
	    &		&		   &	(m) 	     &	  (\AA)      &		\\	
 \hline
M33    & \citet{magrini07}   		& WHT          & 4.2	     &  10				 & $\sim$3-4  & 0.05-0.10 \\
	  & \citet{magrini09b}   	& MMT	  &  6.5           &  5				 & $\sim$10    & 0.02-0.09\\  
	  & \citet{magrini10}   	& MMT	  &  6.5           &  5				 & $\sim$10    & 0.05-0.12\\  
	  & \citet{bresolin10}   		& Subaru	  &8.2  	      & 4-4.5                          & $\ge$10     &0.08-0.12\\  
M31    & \citet{sanders12}		& MMT	  &  6.5           &  5				 & $\sim$10    &0.06-0.2 \\  
	   & \citet{ZB12}		& Keck	  & 10	      & 4.5-5.6			 & $\ge$10      &0.05-0.14\\  
NGC300 & \citet{stasinska13, bresolin09b}		&VLT	  & 8.2	      & 5-10				 & $\le$10	       &0.06-0.15\\			
M81        & \citet{S10}         & MMT	  &  6.5           &  5				 & $\sim$5       &0.10-0.30 \\  
	      &	\citet{S14}           &Gemini-N & 8.2 	      & 5-8				 & $\sim$5       & $<$0.30\\  			 
	      & \citet{patterson12}		& MMT	  &  6.5           &  5				 & $\sim$5       & 0.11-0.26\\  	
\hline
\end{tabular}
\label{tab_obs}
\end{center}
\end{table*}

\section{Migrating populations} 
\label{sec:migrating}

Since  PN progenitors have resided in the galaxy for several Gyr,
it is essential to quantify their  radial migration before
proceeding in the analysis of gradient evolution. To do that, we need
measurements of their velocity along the line-of-sight and proper
motions (the latter being usually not available for extragalactic
PNe). 
Within our galaxy sample, this analysis is possible
only for M31 ad M33, the galaxies with the largest and most complete
sample of PNe with accurate measurements of radial velocities. For
M31, we combine the datasets of \citet{sanders12} and
\citet{Merrett+06}.  

We use only disc PNe, i.e., we remove those PNe
that are classified as H{\sc ii}, halo PNe, or associated to
satellites. \footnote{ \citet{sanders12} have identified halo PNe
  in their sample on the basis of their position along the minor axis
  ($|Y| > 4$ kpc). \citet{Merrett+06} have identified those PNe that
  belong to satellites or background galaxies in their sample on the
  basis of their measured velocities and their proximity to known
  systems (M32, M110, Andromeda IV and VIII, 2MASXi J0039374+420956,
  and MLA93 0953) (see Section 7 of \citet{Merrett+06} for more
  details).} This results in a sample of 731 PNe with measured radial
velocity. For M33, we used the dataset of 140 disc PNe from
\citet{ciardullo04}.

 We adopt two complementary approaches that should
  reveal the effects of different kind of radial motions. In the first
  approach (Section
  \ref{sec:migrating1}), we look for signatures of radial motions in the
  two-dimensional velocity field of the PN population. In the second one (Section \ref{sec:migrating2}), we look for PNe that deviate
  from a simple rotational-disc model, and check if their kinematics is
  consistent with radial motions.  The
  first method is more sensitive to the presence of group of stars
  with an average radial component in their velocity vector; the
  second method is more sensitive to stars with a large radial
  component in their velocity (and small tangential component), and no
  net average radial motion.

\subsection{Signatures of radial motions on the velocity field}
\label{sec:migrating1}

\begin{figure*}
\centering
\includegraphics[width=0.4\textwidth, angle=0]{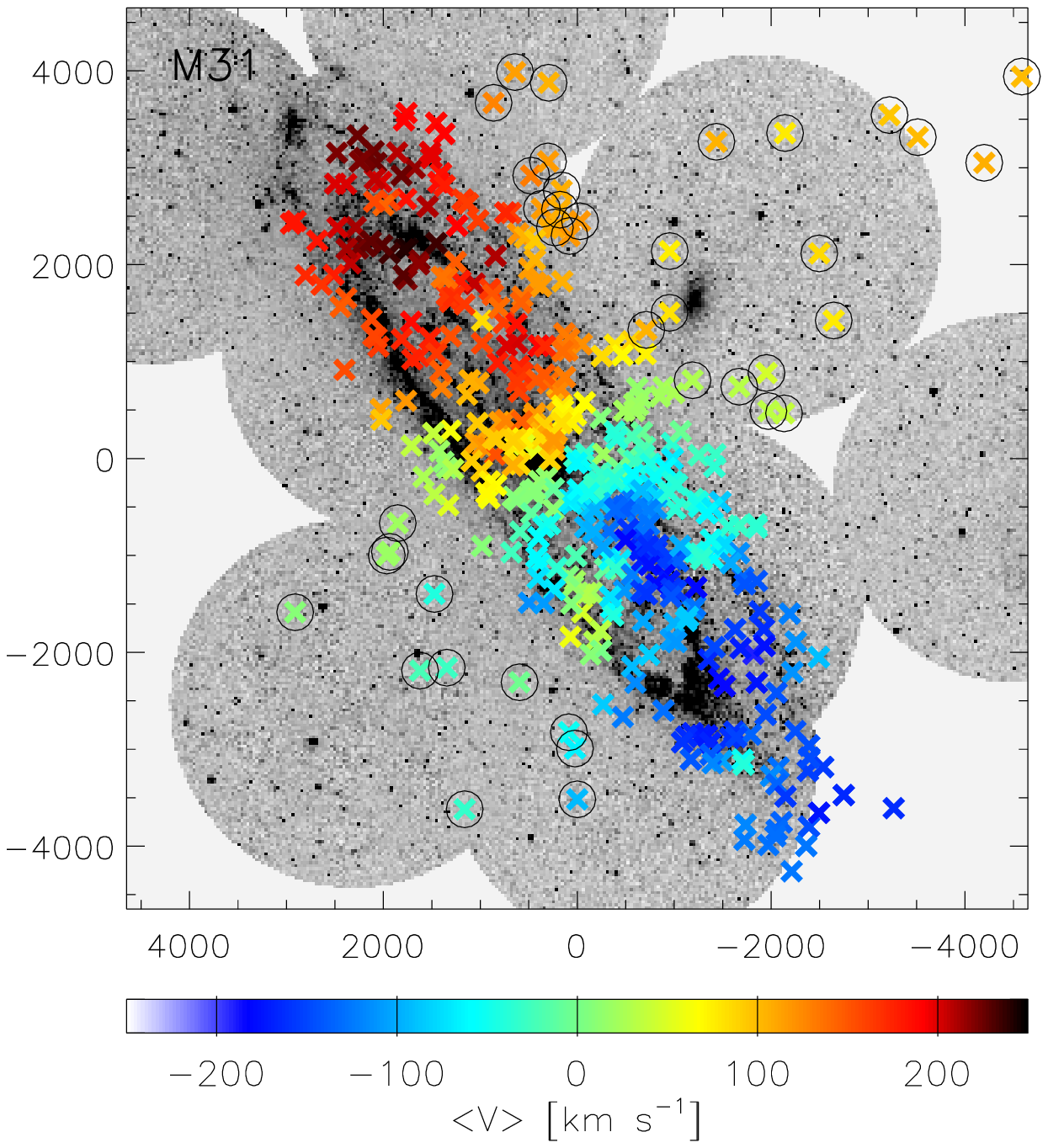}
\includegraphics[width=0.4\textwidth, angle=0]{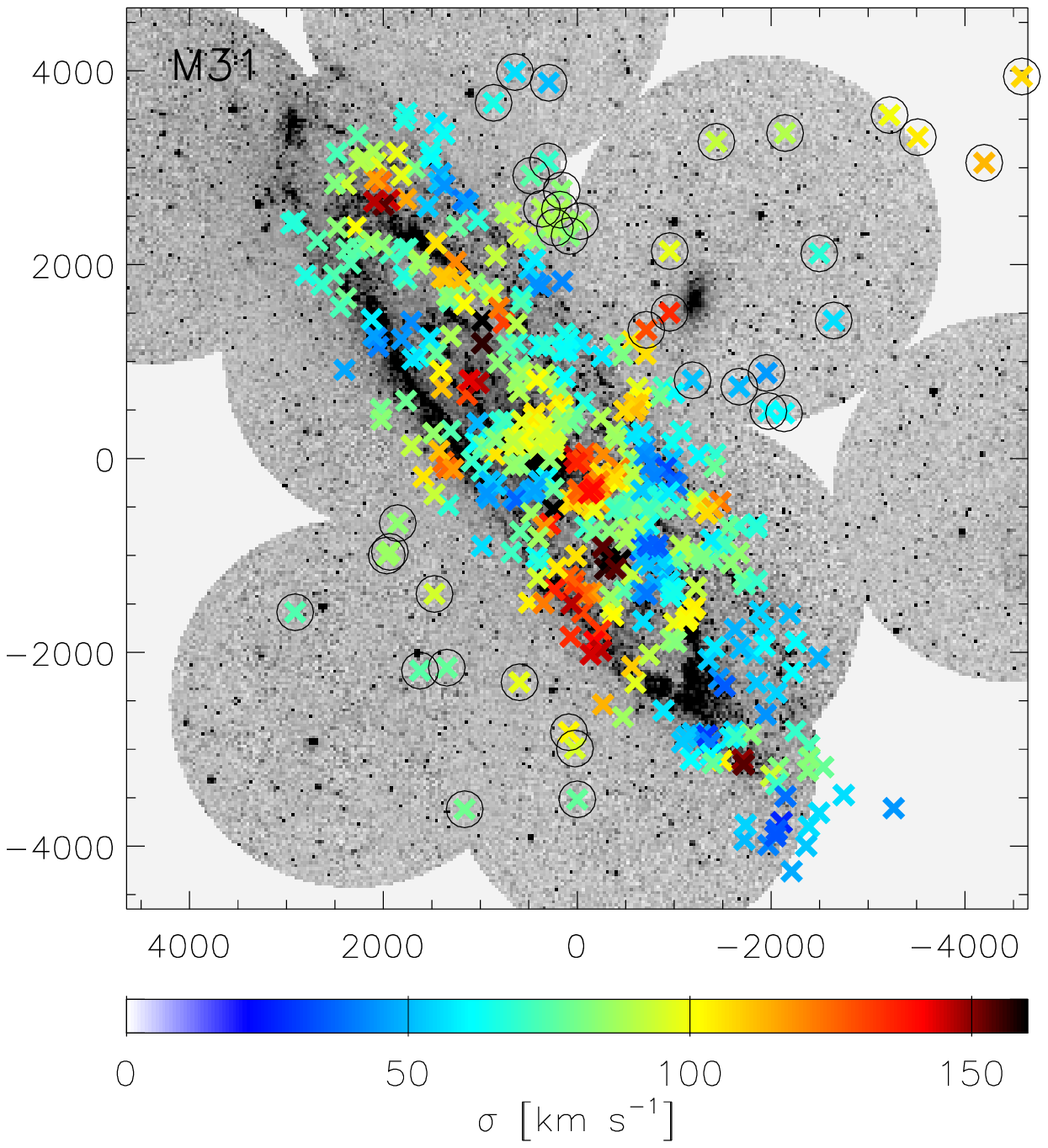}\\
\vspace{0.5cm}
\includegraphics[width=0.45\textwidth, angle=0]{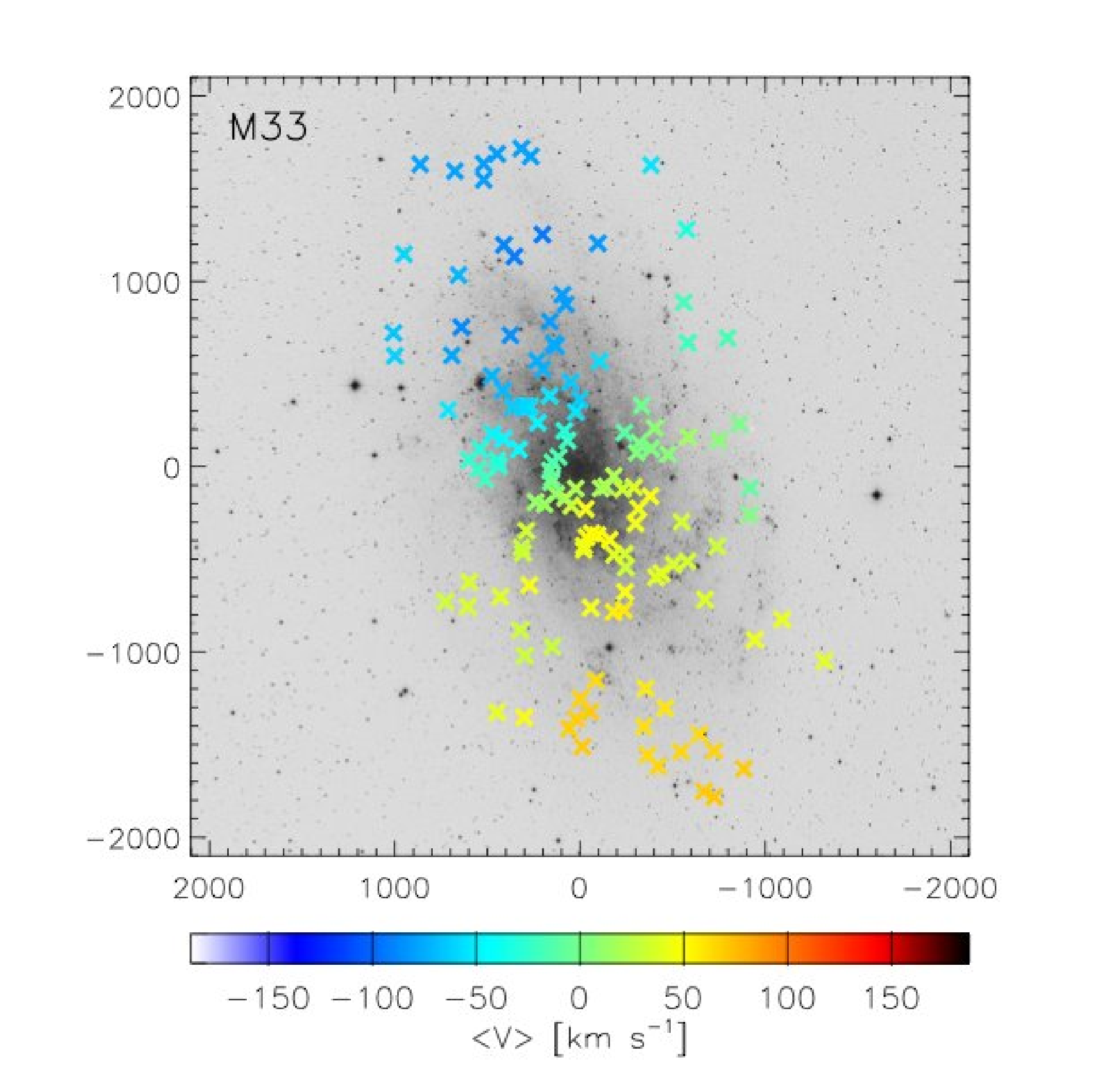} 
\includegraphics[width=0.45\textwidth, angle=0]{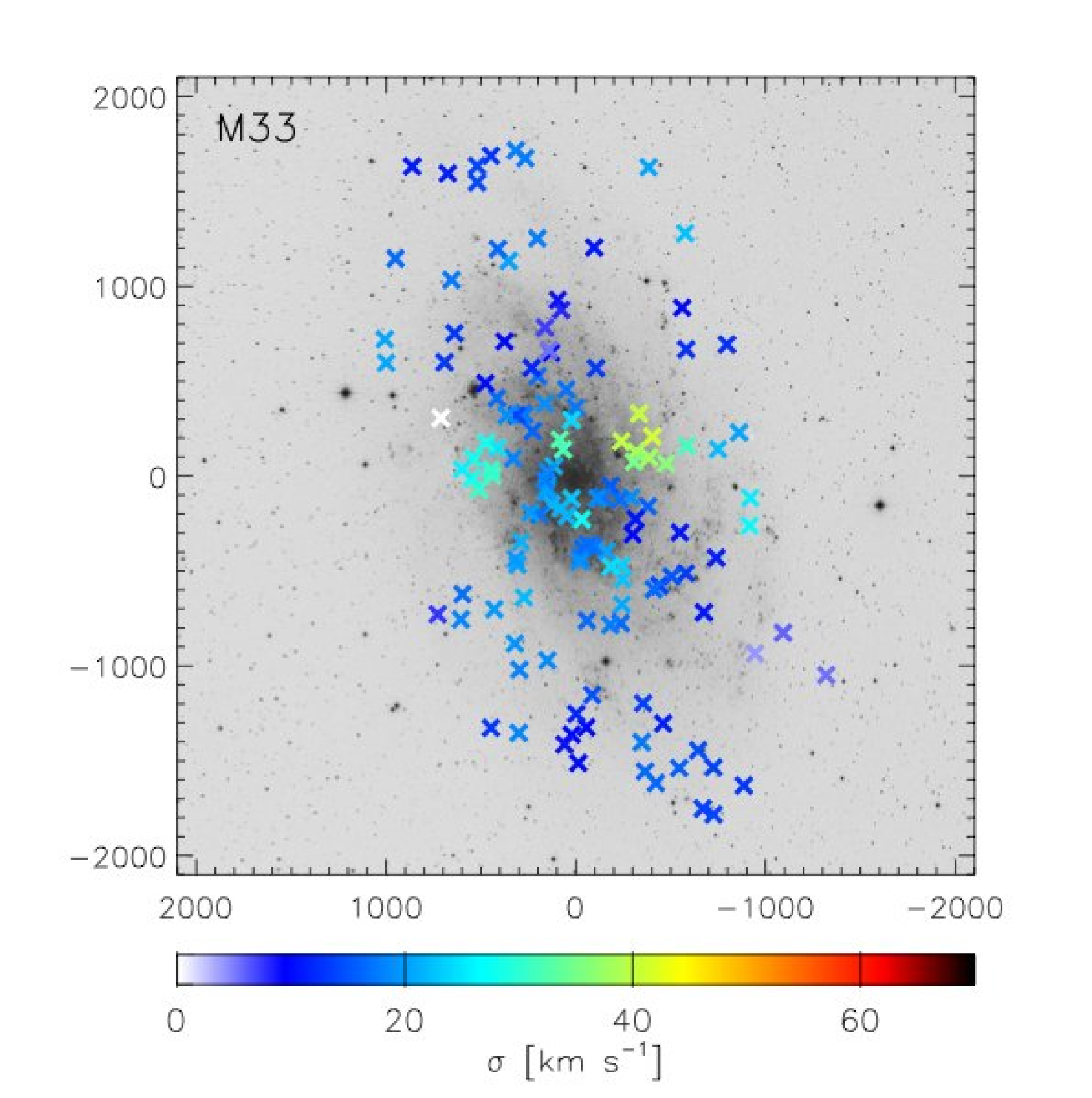}\\
\vspace{0.5cm}
\caption{The reconstructed two-dimensional fields of velocity
(left panel) and velocity dispersion (right panel) for M31 (upper
panels) and M33 (lower panels). Each cross represents the
location of a detected PNe, its colour indicates the value of
$\langle V\rangle$ and $\langle \sigma\rangle$, as computed via the
adaptive Gaussian kernel smoothing in that position. The spatial
scale is given in arcsec. North is up, East is left. PNe in M31 that
are marked with an open circle were excluded from the analysis of
the rotation curve and radial motions.}
\label{fig:vs2d}
\end{figure*}

The two-dimensional fields of velocity and velocity dispersion  of  PNe
are recovered from the observations by smoothing the measured
velocities $v(x,y)$ with an adaptive Gaussian kernel along the spatial
directions \citep{peng04, coccato09}. The size of the kernel
represents a compromise between spatial resolution and number of PNe
within the kernel that are needed to properly reconstruct the local
line-of-sight velocity distribution (LOSVD). The kernel size
automatically adapts according to the local PN number density: regions
with higher concentration of PNe have smaller kernel size, whereas
less populated regions have larger kernel size \citep[see][for
details]{coccato09}.  The two-dimensional
fields  of the reconstructed velocity $\langle V \rangle$
and velocity dispersion $\langle \sigma \rangle $ are shown in Figure~\ref{fig:vs2d}.

Following the formalism by \citet{teuben02}, we model the galaxy
velocity field $\langle V\rangle$ with a simple thin disc model that
accounts for tangential  V$_{\rm rot}$ (i.e., rotation) and expansion motions V$_{\rm exp}$ (i.e.,
radial migration, that can be inward or outward),
\begin{equation}
\langle V(x,y)\rangle = V_{\rm sys} + V_{\rm rot}(R)\cos\theta\sin i+ V_{\rm exp}(R)\sin\theta\sin i,
\label{eqn:model}
\end{equation}
where  V$_{\rm sys}$ is the heliocentric systemic velocity, $(R,\theta)$ are polar coordinates in the plane of the galaxy,
and $(x,y)$ are cartesian coordinates in the plane of the
sky ($\theta=0$ is aligned with the galaxy photometric
  major axis and corresponds to positive velocities along the
  line-of-sight, $\theta$ increases counter-clockwise). The relationship
between the sky and the galaxy planes is given by
\begin{equation}
\tan\theta = \frac{\tan\phi}{\cos i}, \qquad R=r\frac{\cos\phi}{\cos\theta},
\label{eqn:skygal}
\end{equation}
where $(r,\phi)$ are polar coordinates in the plane of the
sky. Inclinations of $i=72^{\circ}$ and $i=55^{\circ}$ were
  used for M31 and M33, respectively.

A positive value of  $V_{\rm exp}$ translates into inward
  or outward motions depending on which side of the galaxy is the
  closest to us. This is  determined by using the differential
  reddening of globular clusters. According to this method, the NW
  side of M31 is the closest to us \citep{Iye+85}, whereas the SE side
  of M33 is the closest to us \citep{Iye+99}. Taking into account our
  sign convention, this leads to have outward motions for positive
  values of $V_{\rm exp}$ in both galaxies.

\begin{figure*}
\centering
\includegraphics[width=0.48\textwidth, angle=0]{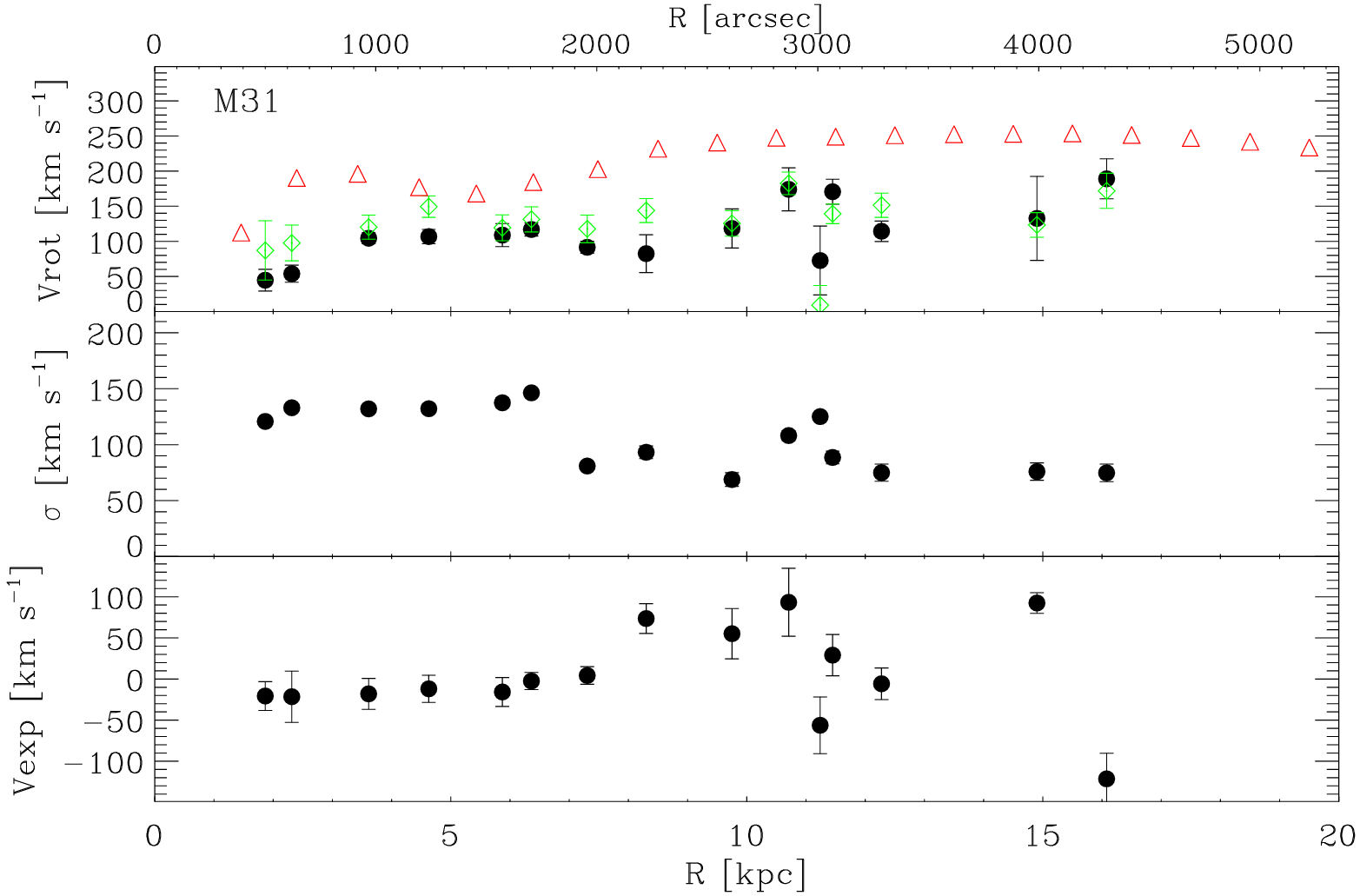}
\includegraphics[width=0.48\textwidth, angle=0]{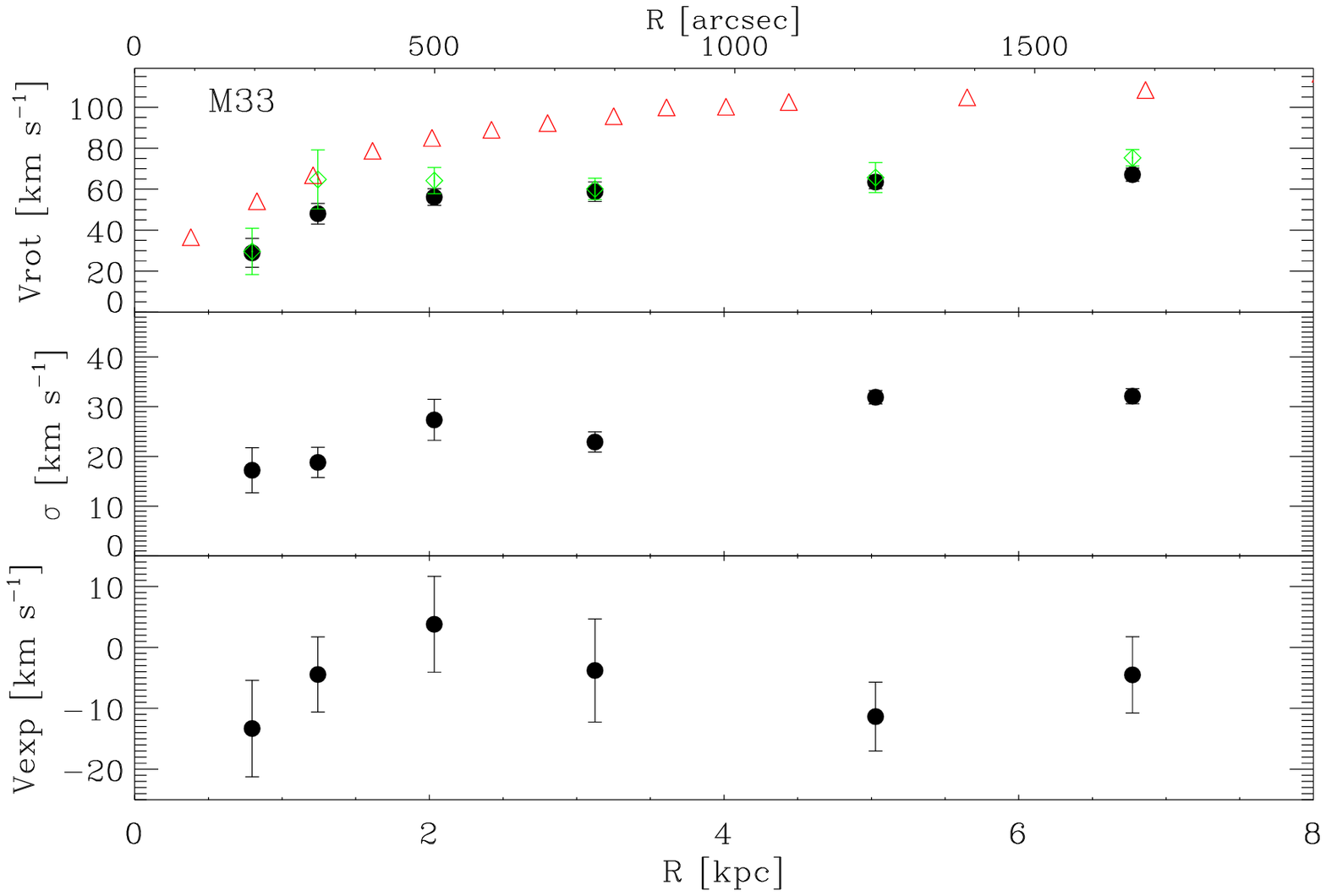}
\caption{Best fit rotation curve (top), velocity
    dispersion profile (middle), and radial motions (bottom) of the PN
    population in M31 (left panel) and M33 (right panel). Black
    symbols show the results of fitting Eq. (\ref{eqn:model});
    green symbols refer to the results when fixing $V_{\rm exp}=0$;
    red triangles show the H{\small I} rotation curve as comparison with M31
    \citep{corbelli10} and  with M33 \citep{Corbelli+00}. Positive values of $V_{\rm exp}$ correspond to outflowing
    motions for both galaxies.}
\label{fig:profiles}
\end{figure*}

We fit eq.~(\ref{eqn:model}) in several elliptical bins to the
reconstructed velocity field computed at the PN position for both
galaxies.  We kept $V_{\rm sys}$ the same for all the
  bins. The boundaries of the bins are spaced according to an inverse
  hyperbolic sine function \citep{lupton98}. The value of $R$ associated to a given bin
  is the median of all the values of $R$ of all the PNe within that
  bin. The best fitting
semi-major axis profiles of $V_{\rm rot}(R)$ and $V_{\rm exp}(R)$ are
shown in Fig.\ref{fig:profiles}. For comparison, we show also the rotation curves of H{\small I} gas for M31
    \citep{corbelli10} and for  M33 \citep{Corbelli+00}. Error bars are computed by means
of Monte Carlo simulations in the following way: we constructed 1000
catalogs of simulated PNe that contain as many PNe as observed and located at the
same position. The velocity of each simulated PN of coordinates
$(x,y)$ is randomly selected from a Gaussian distribution with mean
value $\langle V(x,y)\rangle$, and dispersion $[\Delta
V^2(x,y)+\langle\sigma\rangle^2(x,y)]^{1/2}$, where $\Delta V(x,y)$ is
the measurement error.

The rotation curve of M31 (left panels of Fig.~\ref{fig:profiles}) shows a rising profile reaching 100~km~s$^{-1}$ at 3 kpc (0.22$^{\circ}$) followed by a plateau starting at
$\sim 6$~kpc (0.45$^{\circ}$). At larger radii, the scatter in the
measured velocity increases. This is reflected by the local peaks in
the velocity dispersion maps, and by the presence of kinematic
substructures in the velocity map (upper panels of Fig. \ref{fig:vs2d}). These
features are probably due to the presence of small satellites that
contaminated the PN disc population of M31. The majority of these
features are offset from the galaxy photometric major axis; this
explains the difference between the our PN rotation curve and the one
shown by \citet[their Figure 33]{Merrett+06}, in which only PNe closer
than $0.04^\circ$ to the major axis were considered.  The profile of
radial motions is consistent with 0 km~s$^{-1}$ up to $\sim
7.5$~kpc. For large radii, the contribution of radial motions to the
mean velocity field of M31 becomes relevant, reaching $\sim
100$~km~s$^{-1}$ in modulus, despite the observed scatter. There is
probably a non negligible contribution from small satellites
not identified in \citet{Merrett+06}, but it is not
possible to disentangle migrating PNe from satellite members with
this technique.

In the case of M33, the rotation curve  (right panels of Fig.\ref{fig:profiles}) matches the rotational velocities of the
PNe computed by \citet[][their Figure 7a]{ciardullo04}. It reaches a
maximum amplitude of about $\sim 60$~km~s$^{-1}$ at  $\sim 5$~kpc
and then it remains constant.
The profile of radial motion shows large scatter with a weighted
average of $\langle V_{\rm exp}\rangle = - 6 \pm 7$~km~s$^{-1}$
(about 6~kpc in 1 Gyr). Its radial velocity profile is consistent with 0~km~s$^{-1}$ at
$1\sigma$ level for the majority of the bins (at $3\sigma$ level for
all the bins, except for bins 1 --$\sim 1$~kpc-- and 5 --$\sim
5.6$~kpc--) where there is indication of radial inward motions of
about $-12$~km~s$^{-1}$. 

\smallskip

Our analysis provides strong indication that there is
  a significant $V_{\rm exp}$ component in the PN system of M31. On
the other hand, it provides just a weak indication (at $1\sigma$
level) that  the $V_{\rm exp}$ component is present in
the $1^{\rm st}$ and $5^{\rm th}$ elliptical bins of
M33.  In our simple model, we have directly associated
  $V_{\rm exp}$ to radial motions. However, other deviations from circular motions (such as
  elliptical streaming due to bar-like potential, spiral
  perturbations, warp of the stellar disc) can mimic the presence of
  inward or outwards motions (e.g., \citealt{Wong+04} and references
  therein). Therefore, our measurements of the amplitude of radial
  motions should be considered as upper limits.  In the next
section, we check whether the best fit $V_{\rm rot}$ and $V_{\rm exp}$
are consistent with the individual measurements of PN velocities.

\subsection{PNe that deviate from a pure rotating disc model}
\label{sec:migrating2}

An independent analysis for evaluating radial motion in each
elliptical bin is to count PNe that deviate from a
pure rotating disc model model.  According to Section
\ref{sec:migrating1} we expect to observe deviating PNe outside
7.5~kpc in M31 and on the $1^{\rm st}$ and $5^{\rm th}$ elliptical
bins of M33.  We fit eq.~(\ref{eqn:model}) to the measured PNe
velocities $v(x,y)$ in each elliptical bin setting $V_{\rm exp}=0$. We
then count the number of PNe that deviate more than
$\langle\sigma\rangle(a)$ in each bin, where $\langle\sigma\rangle(a)$
is the mean velocity dispersion in that bin. We then compare this
value with that expected from a pure Gaussian LOSVD. Results are
shown in Tables \ref{tab:outliers} and \ref{tab:outliers2} for M31 and
M33, respectively.

We found that all bins of M31  contain more kinematic outliers than
expected by a Gaussian distribution. The measured
$\langle\sigma\rangle$ represents an upper limit to the velocity
dispersion of the disc PN population, because of the probable presence
of satellites. As a consequence, the estimated number of outliers in
Table \ref{tab:outliers} is a lower limit to the actual value.  

\begin{figure}
\centering
\includegraphics[width=0.5\textwidth, angle=0]{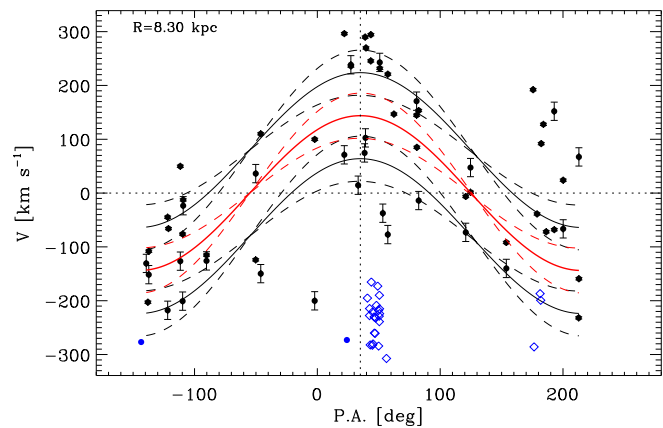}
\caption{Pure rotation disc model fit to PNe velocities in M31 along
  the line-of-sight. The plot shows the fit for bin 8. The filled
  circles are the individual PNe velocities $v(x, y)$ expressed as
  function of the position angle on the sky ($\phi = 0$ in
  eq.~\ref{eqn:model} corresponds to $PA = 35^{\circ}$,
  taken from the HyperLeda database). The red continuous
  curve represents the best fitting rotation model, the red dashed
  lines are the model errors, as computed from the Monte Carlo
  simulations. Continuous black lines indicate the $\pm \sigma$ level,
  with the model errors indicated by black dashed
  lines. Black symbols refer to the analysed PN
    sample, blue symbols locate the discarded PNe: halo PNe as defined
    by \citet[filled blue circles]{sanders12} and satellite PNe as
    identified by \citet[open blue diamonds]{Merrett+06}. }
\label{fig:fit_bin31}
\end{figure}

In
Figure \ref{fig:fit_bin31} we show the azimuthal distribution of the
PNe in M31 for the elliptical bin number 8 (the one with the largest
number of outliers compared to the expected
number).  The corresponding figures for the other bins
  are shown in Appendix A.  The Figure shows that there are plenty of
PNe whose line-of-sight velocity is consistent with being on radial
migration. They are PNe with nearly 0 velocity, located at
the position angles where the rotation is maximum, and  PNe
with high velocity located at the position angles where the rotation
is 0. Indeed, from eq.~(\ref{eqn:model}) we expect the radial motions
to be minimal (maximal) in correspondence of the position angles where
the rotation is maximal (minimal).  

 In the same
  Figure, we also show the location of PNe classified as halo or
  members of satellite systems (which are excluded from the fit). The
  majority of these PNe are clumped and well separated from the PNe
  associated to M31. However, there is some overlap between these
  systems, which can be responsible for some of the radial
  motions. Also, it is worth noticing that some of the PNe associated
  to M31 and that deviate from the circular model are clumped
  together. This indicates that they are kinematically distinct from
  the PNe in the disc; whether this kinematic separation is entirely
  due to radial motions or to being part of an (unknown) satellite is
  beyond the scope of this paper.

\begin{figure*}
\centering
\includegraphics[width=1.0\textwidth, angle=0]{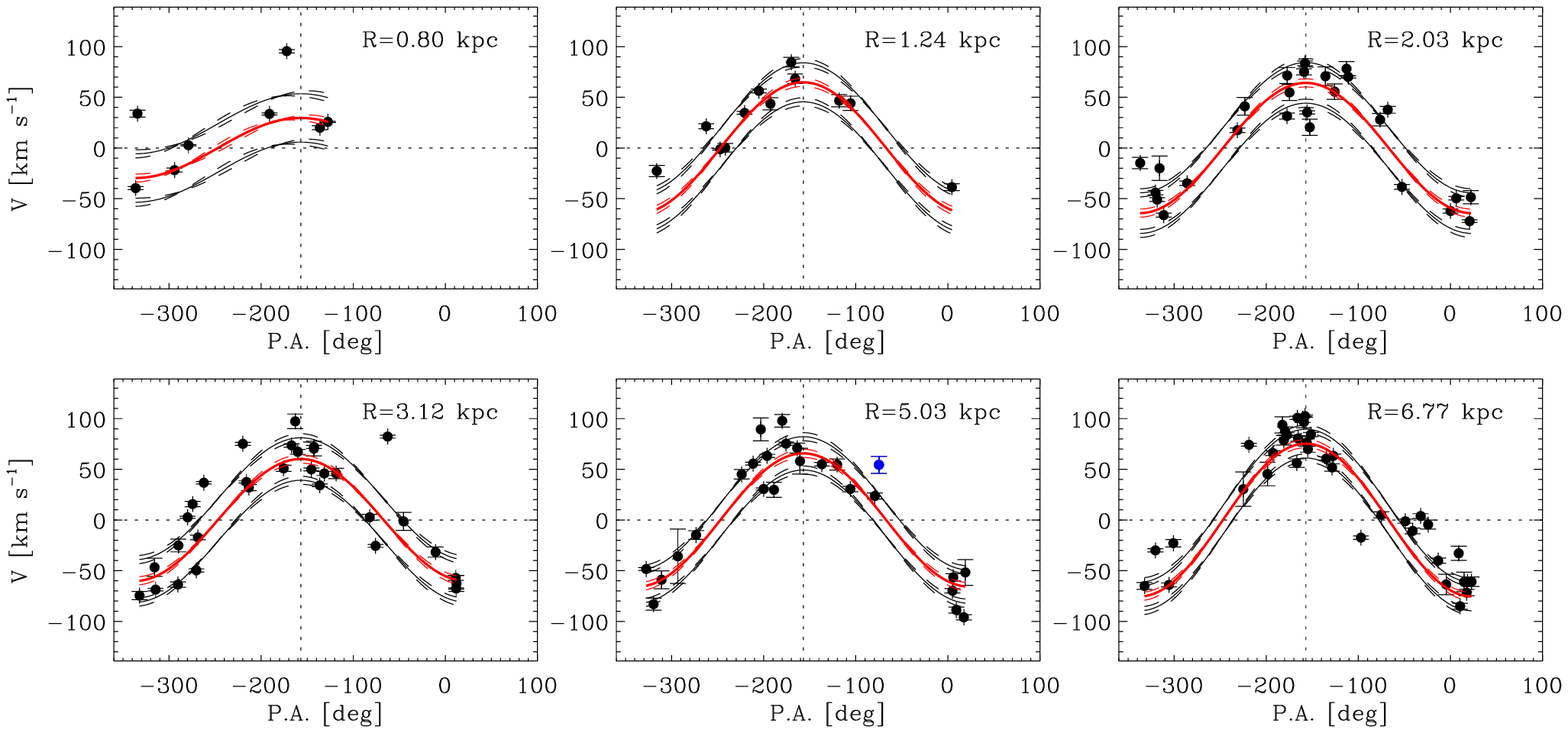}
\caption{Same as Figure \ref{fig:fit_bin31}, but for all the
  elliptical bins of M33. The term $\phi = 0$ in Eq.~\ref{eqn:model} corresponds to $PA =
  -153^{\circ}$ (taken from the HyperLeda database). }
\label{fig:fit_bin}
\end{figure*}

In M33,  in contrast with Section \ref{sec:migrating1}, the $1^{\rm
st}$ bin does not show more outliers than what expected from a
Gaussian distribution. Bin 5 has 9 outliers, whereas 8 are
expected. Of those outliers, 8 are located along the photometric
major axis, where PNe in radial orbits should have zero velocity
projected along the line of sight (LOS, see Eq. (\ref{eqn:model})). However, those 8 PNe have velocity much higher (in
absolute value) than the systemic velocity. Therefore their
kinematics is not consistent with radial motions. One PN (in blue in
Figure \ref{fig:fit_bin}) is located along the galaxy photometric
minor axis. This is the direction where PNe in radial motions have
LOS velocity higher than the systemic velocity.  The positive sign
of the velocity of that PN is consistent with outward radial
motions. However, this is in contrast with what found in section
\ref{sec:migrating1}: radial motions in bin 5, if present, should be
directed inwards.  Similar arguments hold for bin 4, where an
outlier (outside $3 \sigma$) is observed along the galaxy
photometric minor axis and has a velocity $75$~km~s$^{-1}$,
consistent with outward motions.  However, the indication for bin
4$^{\rm th}$ -- if we ignore the error bar -- is that motions are
directed inwards.

The two different approaches give weak indications of the presence of
radial motions in the PN population of M33, which are in contrast with
each other.  Even if we consider the upper limits of the radial
motions found in our analysis, we can conclude that they do not imply
a significant variation for the metallicity gradients of the PN
population.  In particular, a slight flattening could be produced by a global outward motion, for which we estimate an upper limit of $\sim 1$~km
  s$^{-1}$ \footnote{The upper limit is estimated from $<$V$_{\rm exp}$$>$=(-6$\pm$ 7)~km
  s$^{-1}$}.    This might be a significant
amount, if integrated along the whole life-time of a PN progenitor. However, the
amount of PNe presently affected by radial migration does not exceed 7\%, as
estimated by the number of outliers in Table~\ref{tab:outliers}.  The
dramatic change in the slope of metallicity gradient foreseen by
simulations \citep[e.g., ][]{roskar08} can be obtained assuming that
the migrating stars contribute to $\approx 50$\% of the total mass. 
These numbers are, however, not easy to compare with observations 
  because our measurements describe possible migration at the current
  time, whereas the quantity in the simulations refer to the galaxy
  lifetime, and many stars end up on circular orbits after migration.
On the other hand, the effect of radial migration is more visible in
the scatter of measurements rather that the slope itself
\citep{Grand+15}.

In conclusion, our analysis supports the existence of significant
radial motions in the PN system of M31.  The present-time migration
as measured from PNe and, especially, the considerable component of
migrating PNe, might be compatible and responsible of the strong
variation of the slope of the gradient, from that of PNe to that of H{\sc ii} regions
(see Table~\ref{tab_meanabu}).  On the other hand, our analysis does
not support the existence of significant radial motions in the PNe
system of M33 at present time, with a  negligible 
 effect  compared to the uncertainties.

\begin{table}
\caption{{Number of PNe in M31 that deviate more than 1$\sigma$ from a pure
rotating thin disc model compared with that expected from a pure
Gaussian distribution for the elliptical bins. The semi-major axis
of each bin is indicated in the second column.}}
\centering
\begin{tabular}{l c c c}
\hline
bin  & R   &  Outside 1$\sigma$ & Expected \\
     & (kpc)  &                    &          \\
\hline
 1 &  1.87&   4 &  4.0 \\ 
 2 &  2.31&   23& 13.0 \\
 3 &  3.61&   26& 21.1 \\
 4 &  4.63&   28& 24.1 \\
 5 &  5.87&   37& 24.4 \\
 6 &  6.36&   26& 21.1 \\
 7 &  7.30&   20& 18.5 \\
 8 &  8.30&   34& 20.5 \\
 9 &  9.75&   30& 20.5 \\
 10& 10.71&   24& 19.5 \\
 11& 11.24&   37& 31.0 \\
 12& 11.45&   21& 10.2 \\
 13& 12.28&   15&  7.9 \\
 14& 14.90&   7 &  4.0 \\
 15& 16.08&   3 &  1.7 \\
\hline
\end{tabular}
\label{tab:outliers}
\end{table}
       
\begin{table}
\caption{Same as Table~\ref{tab:outliers} for M33.}
\centering
\begin{tabular}{l c c c}
\hline
bin  & R   &  Outside 1$\sigma$ & Expected \\
     & (kpc)  &                    &          \\
\hline
 1   & 0.80  &     2              &    2.6  \\
 2   & 1.24  &     3              &    4.0  \\
 3   & 2.03  &     9              &    8.6  \\
 4   & 3.12  &    10              &   10.2  \\
 5   & 5.03  &     9              &    8.3  \\
 6   & 6.77  &    12              &   12.5  \\
\hline
\end{tabular}
\label{tab:outliers2}
\end{table}

\section{The time-evolution of radial metallicity gradients} 
\label{sec4}
 
In order to determine time-evolution of radial metallicity gradients we proceed with the  homogenisation of the literature samples in several ways: 

{\em 1. Levelling out the radial scale.}  Spiral galaxies have
different dimensions, often related also to the morphological type,
with Sa galaxies being more extended that Sd ones.  Several studies in
the past have shown that the radial metallicity gradients can be
related to several galactic properties, such as their morphology
\citep[e.g.,][]{mccall85, VE92}, mass \citep[e.g.,][]{zaritsky94,
  MR94}, presence of a bar \citep[e.g.,][]{zaritsky94, roy96},  and
interaction stage \citep[e.g.,][]{rupke10}.  It has been
claimed that the dependence on galaxy properties is evident only if
the gradients are expressed in dex~kpc$^{-1}$ and it disappears when
expressed in unit of effective or optical radius.  To make sure that
the different scale lengths of galactic radii do not produce
insignificant correlations with other galactic properties, a good
approach is to use, e.g., the optical radius $R_{25}$ (the 25
  mag arcsec$^{-2}$ isophotal ratio), the effective or
  half-light radius  $R_{e}$, or the disc scale-length (the scale-length of
  the exponential describing the profile of the disc), $R_d$ .  To this
respect, it is worth noting that the works by \citet{sanchez14},
\citet{ho15}, \citet{pilyugin14}, and \citep{BK15}, where radial metallicity gradient slopes are expressed in terms of 
a reference radius,  levelled the differences across
galaxies. 

 The typical gradients of these works range from $-0.12\,{\rm dex}/(R/R_{e}$) for \citet{sanchez14}, to 
$-0.39 \pm0.18\,{\rm dex}/(R/R_{25}$) for  \citet{ho15}, $-0.32\pm 0.20\,{\rm dex}/(R/R_{25}$) for \citet{pilyugin14}, and 
$-0.045\pm 0.013\,{\rm dex}/(R/R_{d}$) (O3N2 calibrator) for \citet{BK15}. 
 Note that \citet{sanchez14}, 
\citet{ho15} and  \citet{BK15} express the gradients in different units: for instance, the \citet{sanchez14}'s
gradient would be $-0.20\pm 0.18\, {\rm dex}/(R/R_{25}$) if expressed
in a $R_{25}$ scale,  as done by \citet{ho15} assuming an
  exponential disc with the typical central surface brightness for
  normal spirals and the empirical conversion between the different
  metallicity calibrations \citep{KE08}.

  While all the four examples mentioned above are based on
strong-line abundances, there is purpose in comparing the gradients
of galaxies of different morphological types based on a common
physical scale, as, e.g., $R_{25}$. In this way we can also directly
compare same-population gradients of different galaxies.


{\em 2. Binning the data.}  Smoothing the data gives in general a
better idea of their behaviour.  The binning procedure in the analysis
of the radial metallicity gradients was performed, e.g., by
\citet{MQ99} in the study of the gradient of PNe and by
\citet{huang15} in the study of the red clump star gradient. The
procedure gives more robustness to the data, removing possible local
discrepancies and uncertainties, and defines an average abundance at
each galactocentric distance.  In addition, it avoids to over-weight
some regions in the parameter space where, for several reasons, we
might have more data.  Each bin is represented by the average
abundances of all objects having distances between $\pm 0.05 R/R_{25}$
from the central value.  For each galaxy, the different
  literature sources give comparable errors, thus each data point was
  equally weighted to produce the final bin value. The error is the
standard deviation of the mean.

{\em 3. Comparing common radial bins.}
In the CALIFA sample the gradients of many galaxies
show evidence of a flattening beyond $\sim$2 disc effective radii.
This is consistent with several spectroscopic studies of external
galaxies \citep[e.g.,][]{bresolin09, bresolin12} and of our own
Galaxy \citep[e.g.,][]{magrini09, lepine11, huang15}.  So, the
comparison of different radial regions can lead to misleading
results,  especially if some of them include very external part of the
galactic discs.  In the following, we compare similar radial
ranges for the H{\sc ii} region and PN population in each galaxy, excluding
bins outside $\sim R_{25}$ where possible changes of slopes might
occur.

\section{Results}
\label{sec5}

\subsection{Individual galaxies}

\begin{figure}
\centering
\includegraphics[width=0.35\textwidth, angle=270]{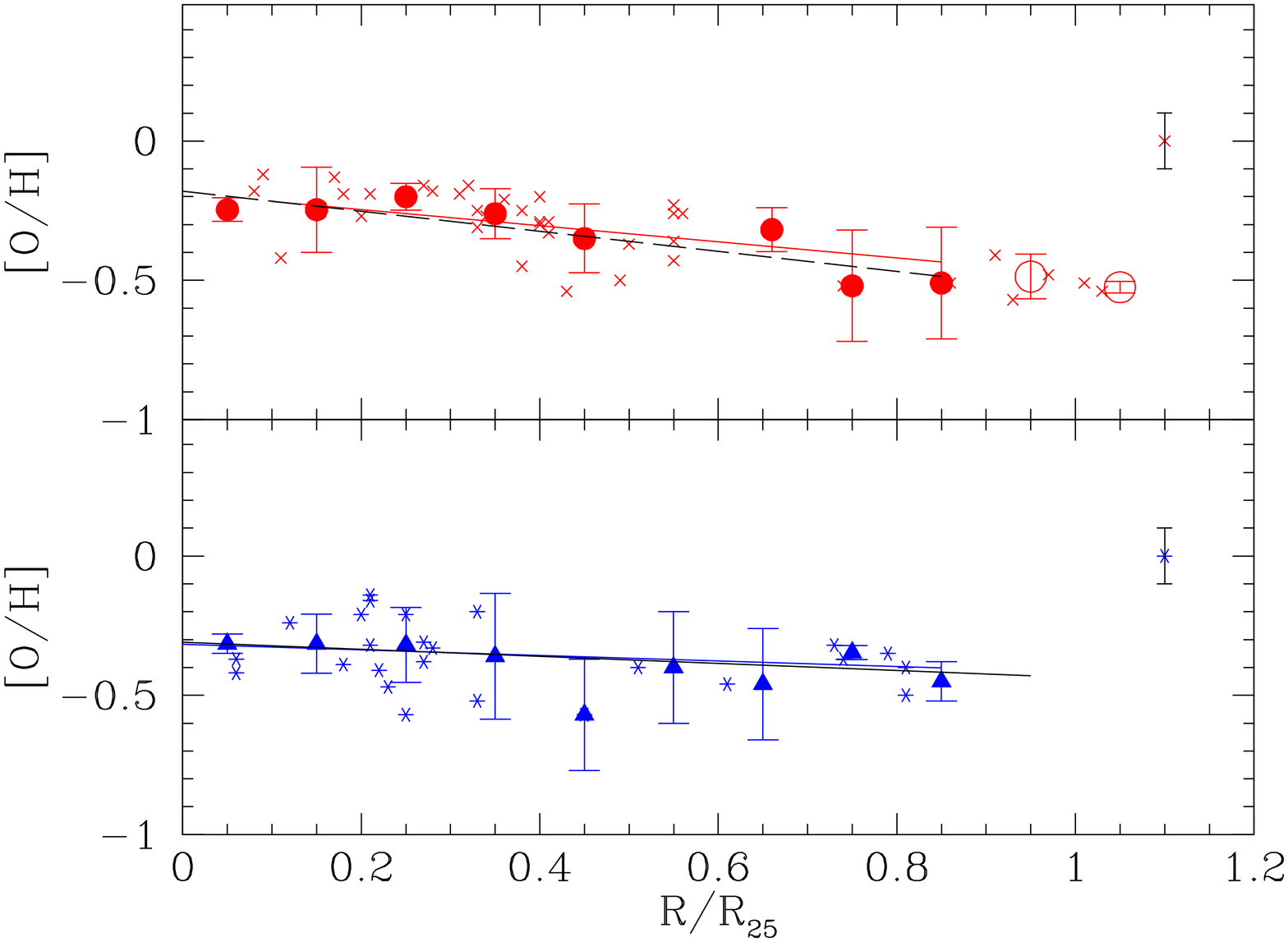}
\includegraphics[width=0.35\textwidth, angle=270]{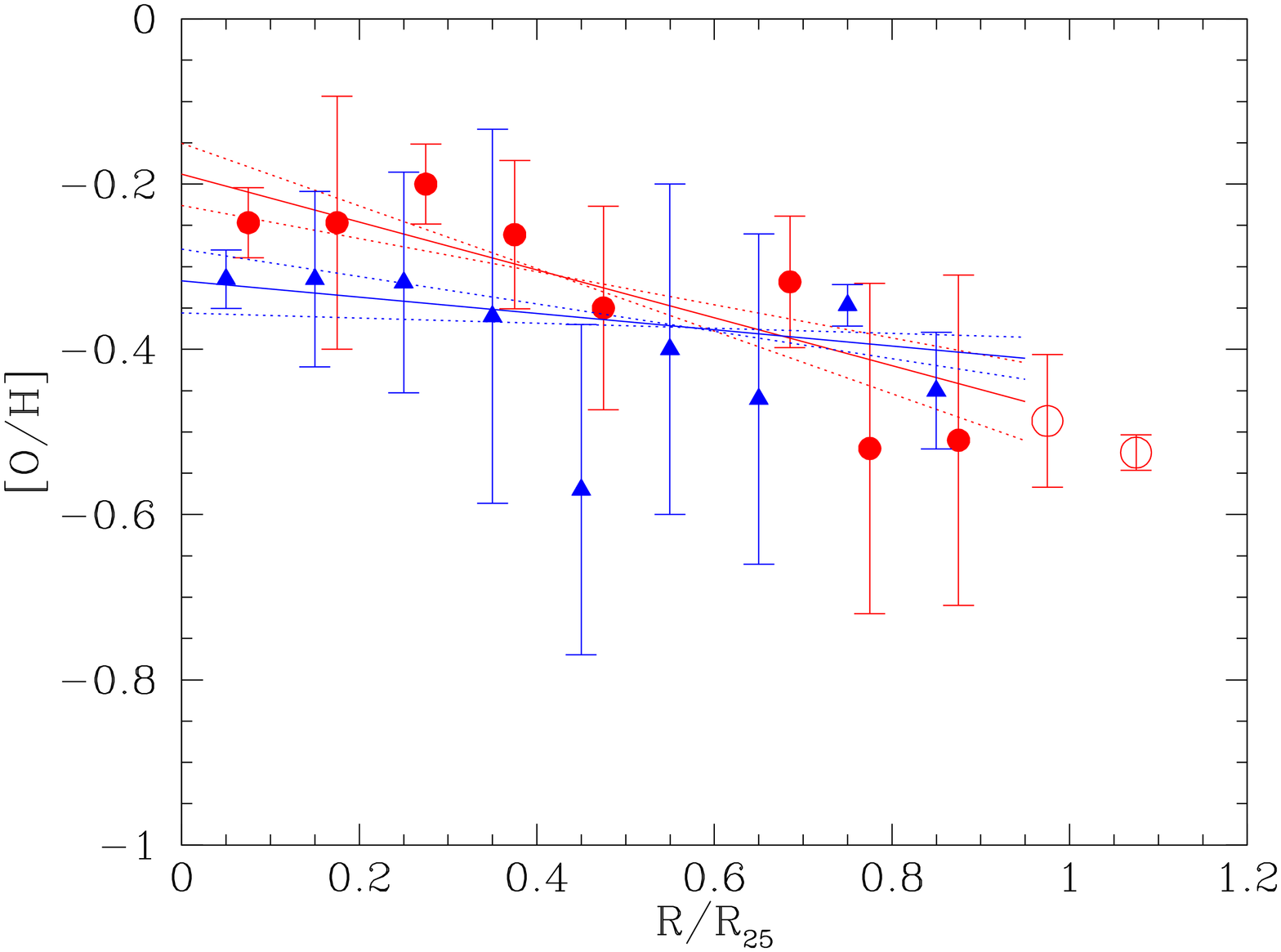}
\caption{NGC300. {\em Upper panel}:  individual and binned results for
H{\sc ii} regions (red circles and crosses for binned and individual results, respectively) and PNe (blue triangles and crosses for binned and individual results, respectively) in NGC300.  The
O abundances of PNe and of H{\sc ii} regions are from \citet{stasinska13}
and \citet{bresolin09b}. The  dashed black lines are the gradients
of \citet{stasinska13}. 
The  continuous  lines are weighted linear fits computed in the radial regions where both
populations are available.  The typical errors in O/H for each dataset are shown on the right side of each panel. {\em Lower panel}: weighted
linear fits of PN and H{\sc ii} regions binned metallicities as in the upper panel. Each point
corresponds to the average metallicity of the PN or H{\sc ii} region
population in a radial bin of $0.1 R/R_{25}$.  Empty symbols are
for radial positions outside $R_{25}$ or for populations that 
have no correspondence in the other sample (excluded from the fits). 
Dotted lines indicate the
errors on the slopes and on the intercepts.}
\label{NGC300}
\end{figure}

\begin{figure}
\centering
\includegraphics[width=0.35\textwidth, angle=270]{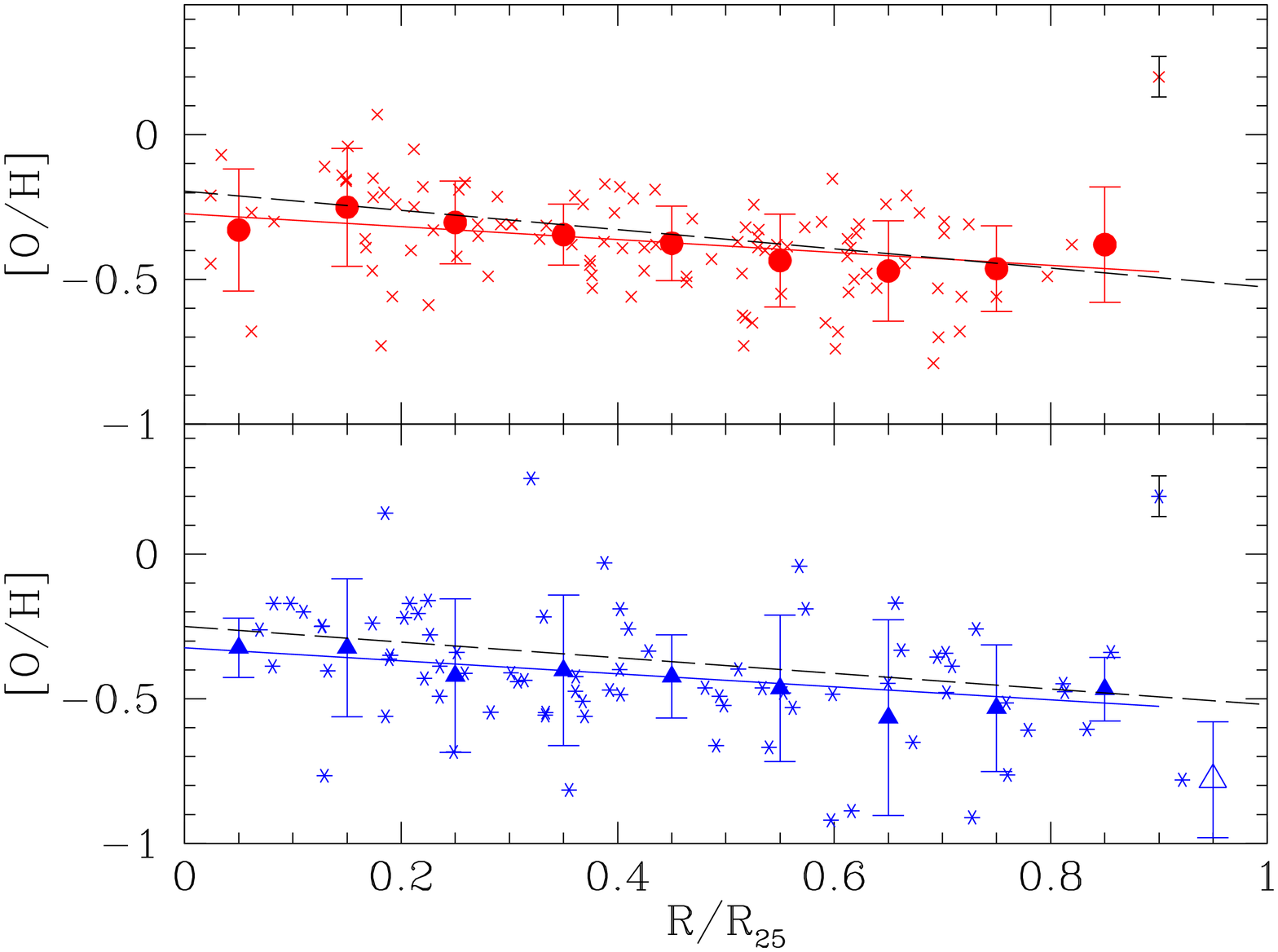}
\includegraphics[width=0.35\textwidth, angle=270]{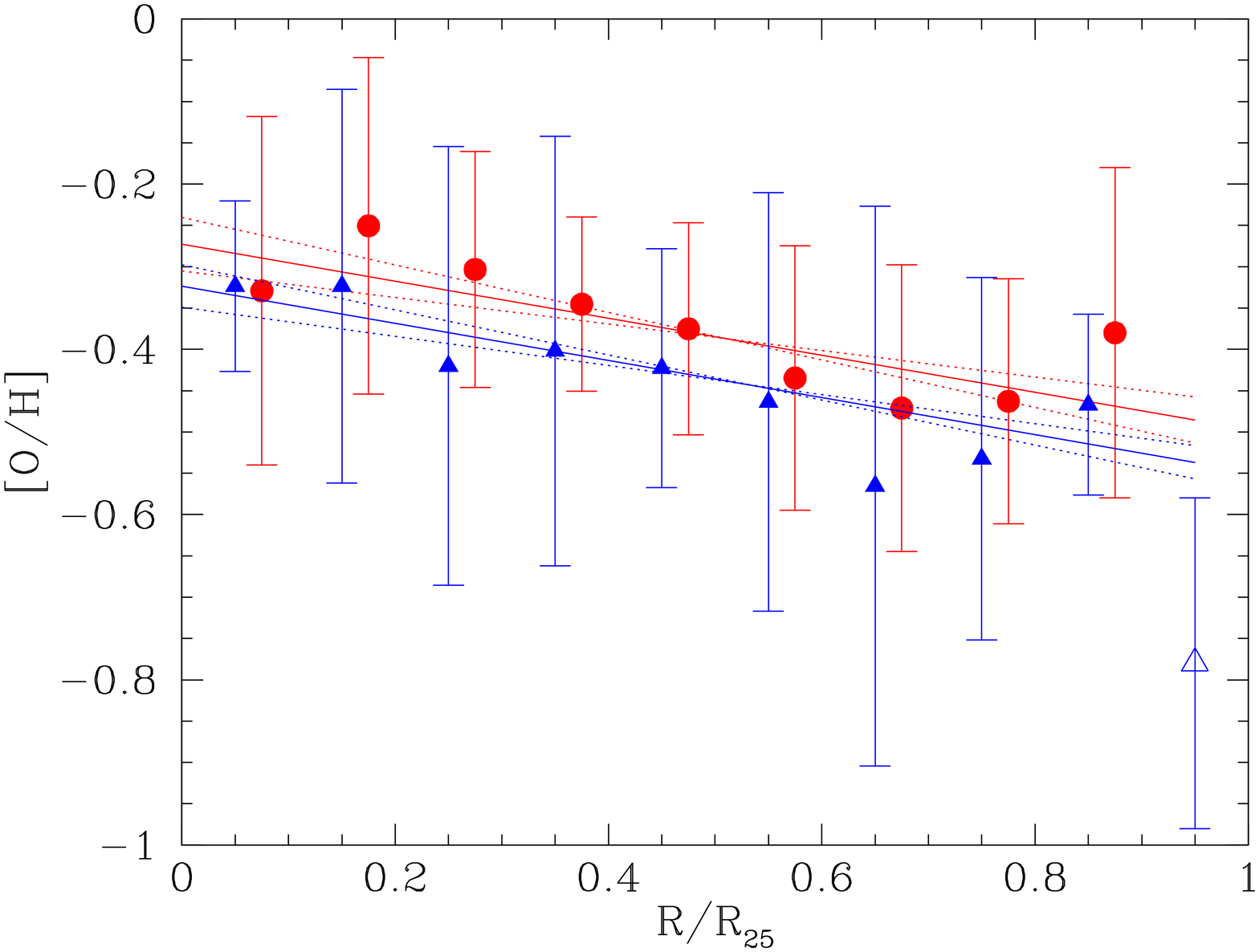}
\caption{M33. {\em Upper panel}:  individual and binned results for H{\sc ii}
regions (upper panel) and PNe (lower panel) in M33.   The O
abundances of PNe are from \citet{magrini09b} and \citet{bresolin10} (blue
stars) and of H{\sc ii} regions from \citet{magrini07,magrini10} and
references therein, and \citet{bresolin10}  (red crosses). The
continuous black lines are the gradient of \citet{magrini10} for
H{\sc ii} regions (their whole sample) and the non-Type I PN sample of
\citet{magrini09b}, in the upper and lower panels, respectively.
{\em Lower panel}: weighted linear fits of PN and H{\sc ii} region binned
metallicities.   Symbols and curves as in Fig.~\ref{NGC300}.  }
\label{M33}
\end{figure}

\begin{figure}
\centering
\includegraphics[width=0.35\textwidth, angle=270]{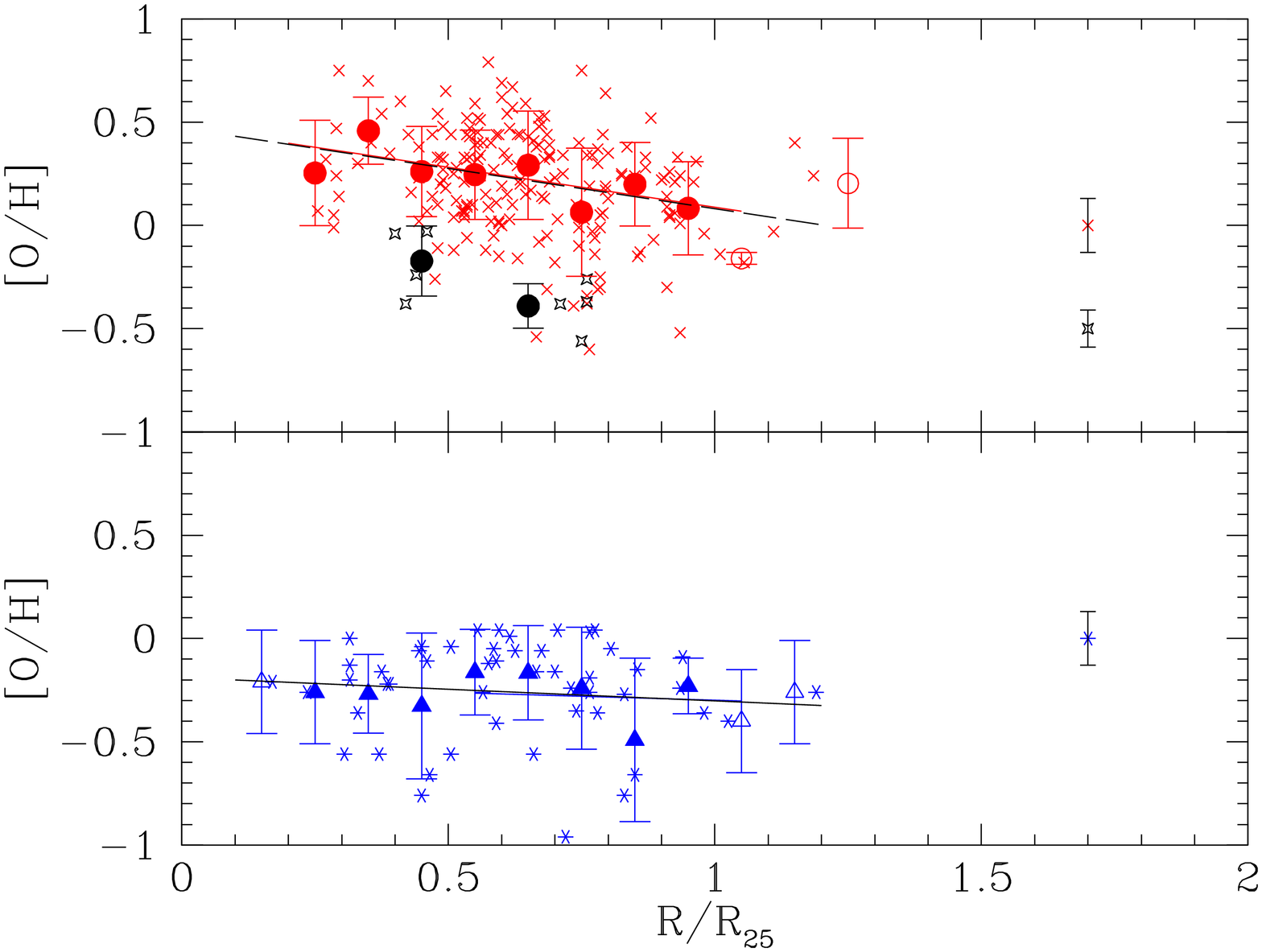}
\includegraphics[width=0.35\textwidth, angle=270]{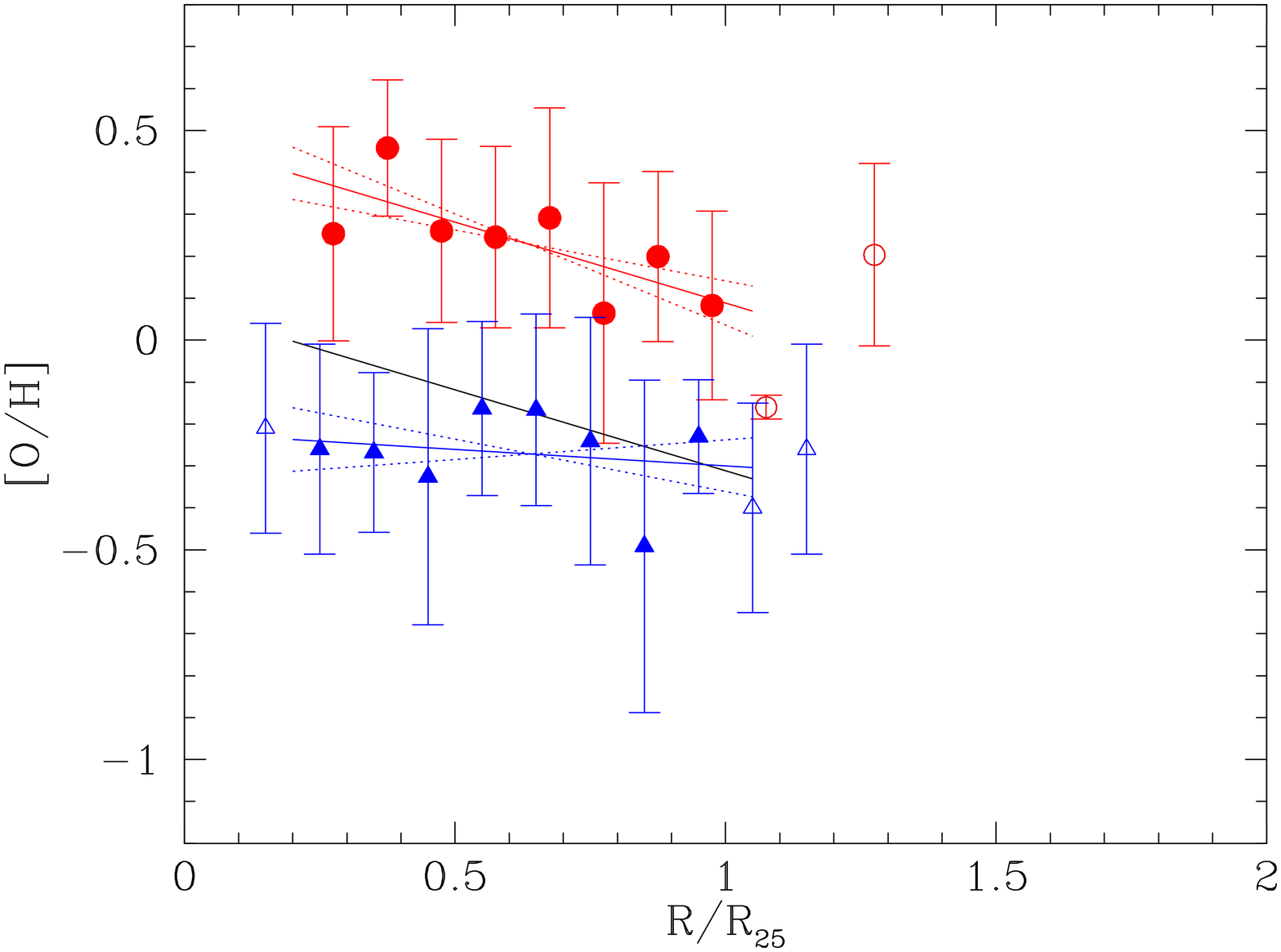}
\caption{M31. {\em Upper panel}: individual and binned results for H{\sc ii}
regions (upper panel) and PNe (lower panel) in M31.  The O
abundances of PNe are from \citet{sanders12} (blue stars) and of
H{\sc ii} regions from \citet{ZB12} (black empty stars) and from
\citet{sanders12} (red crosses). The black filled circles are the H{\sc ii} regions binned metallicities of \citet{ZB12}. 
{\em Lower panel}: weighted linear fits of PN and H{\sc ii} region binned metallicities. Symbols and curves as in Fig.~\ref{NGC300}.  }
\label{M31}
\end{figure}

\begin{figure}
\centering
\includegraphics[width=0.35\textwidth, angle=270]{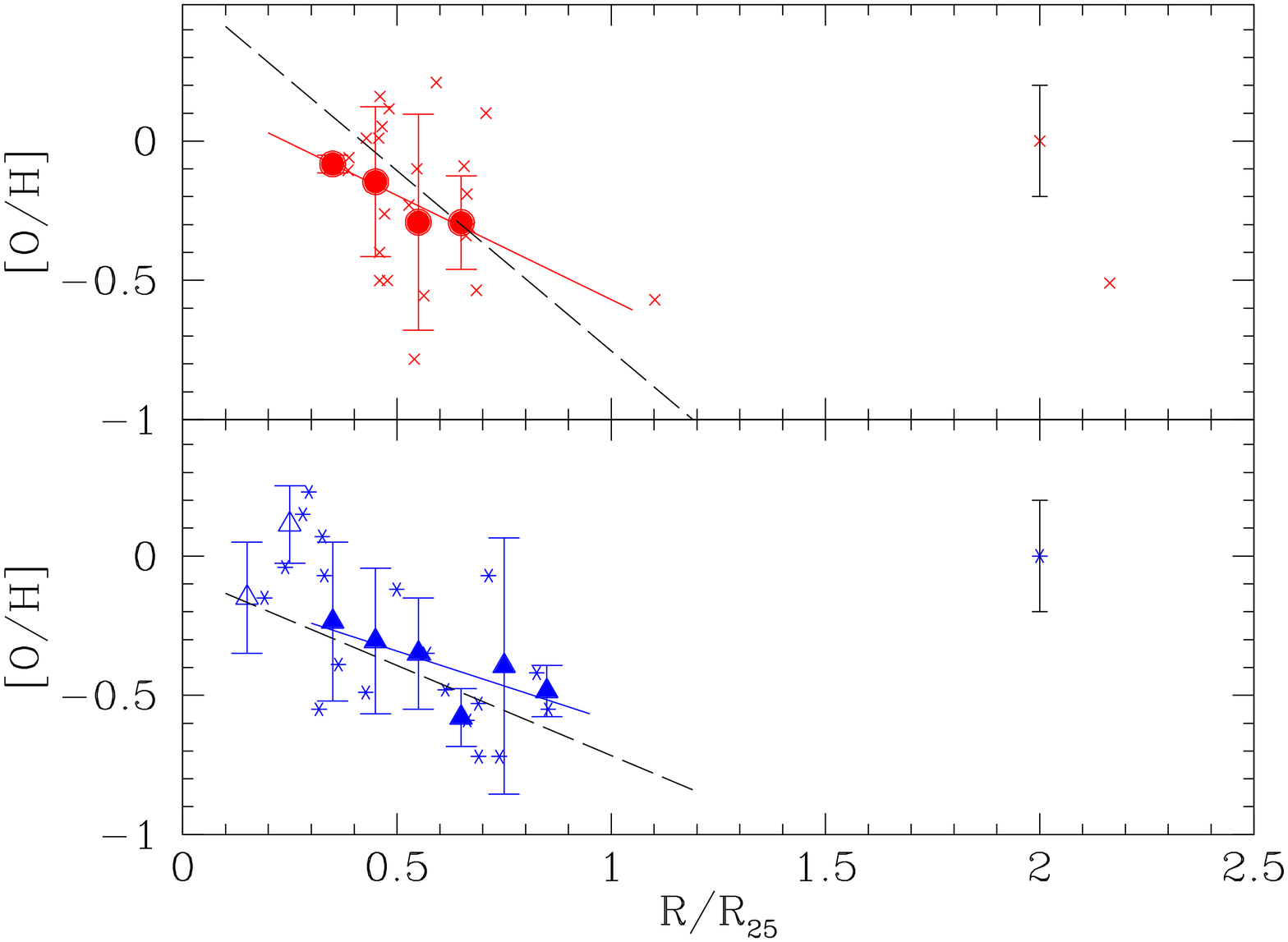}
\includegraphics[width=0.35\textwidth, angle=270]{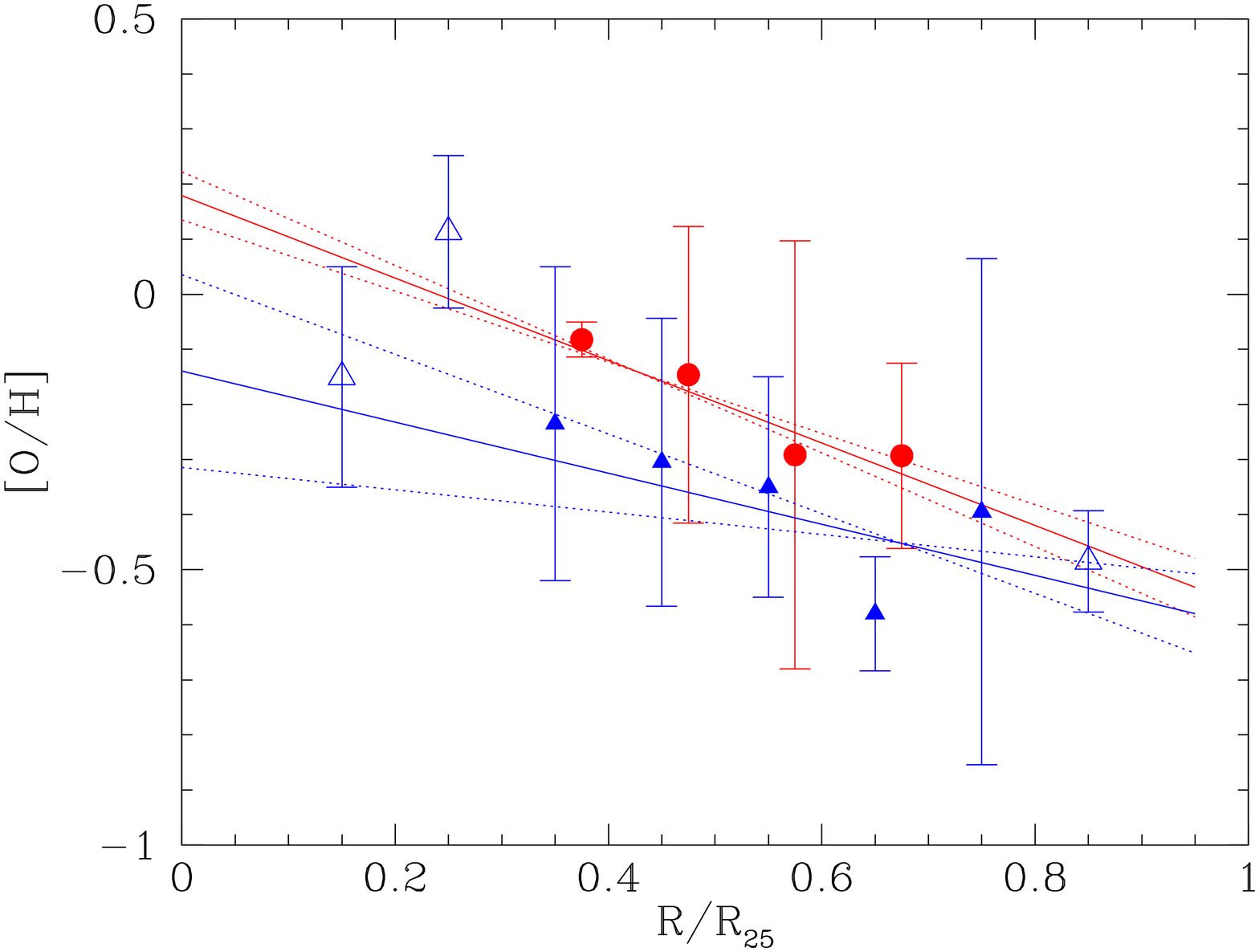}
\caption{M81. {\em Upper panel}:  individual and binned results for H{\sc ii}
regions (upper panel) and PNe (lower panel).   The O abundances
of PNe are taken from \citet{S10} and those of H{\sc ii} regions  from \citet{S10,S14} and 
\citet{patterson12}. The
continuous black lines are the gradient of \citet{S14} for PNe and that of \citet{S14}
for H{\sc ii} regions.  Lower
panel: weighted linear fits of PN and H{\sc ii} region binned metallicities.
Symbols and curves as in Fig.~\ref{NGC300}.}
\label{M81}
\end{figure}

In Figs.~\ref{NGC300}, \ref{M33}, \ref{M31}, and \ref{M81} we show
the results  of our analysis.  In the upper
panels of each figure, we show the individual measurements, 
the gradients obtained from the linear weighted fit of binned data  and the original literature gradients.
In the bottom panels, we present the binned abundances with their linear weighted fits.

{\em NGC300.} Fig.~\ref{NGC300}  shows the results  for NGC300. The two outermost H{\sc ii} region  bins have not been included in the fit, 
since there are no observed PNe in the corresponding bins.
Their exclusion explains the small
difference with the gradients derived by \citet{stasinska13}.  As
already noticed by \citet{stasinska13}, there is a clear difference
in the slopes of the H{\sc ii} region and PN gradients: the gradient  in
H{\sc ii} regions is steeper than the PN gradient. 
For this galaxy we cannot evaluate  the effect of radial migration, thus the measured steepening 
with time is an upper limit: the gradient of PNe could have been steeper than the 
observed one.

{\em M33.} Fig.~\ref{M33}  shows the results for M33. In this case we
excluded on the outermost bin in the PN sample.  As already noticed
by \citet{magrini10}, the PN and H~{\sc ii} region gradients are indistinguishable within the errors,
and  present only an offset in metallicity. For M33 the effect of radial migration 
is negligible, and thus the comparison between the two populations shows the
``true'' evolution of the gradient, which is null or extremely small. 

{\em M31.} The gradients of M31 are shown in Fig.~\ref{M31}. In the first
panel,  we show also the {\em direct}-method abundances
of \citet{ZB12} for a small number of H{\sc ii} regions. These abundances
are much lower that those derived by \citet{sanders12} with the N2
indicator. This is a well-known systematic effect due to strong-line
	vs. {\em direct} methods \citep[e.g.,][]{KD02}.  However, the slopes of the gradient in the radial region where
  the two methods overlap are in rough agreement with the gradient
  from the {\em direct} method, the latter being slightly steeper.  As shown by
  \citet{S15}, the strong-line abundances can indeed give ''first
  order'' constraints to galactic evolutionary models, even if more
  uncertain than the {\em direct}-method abundances.  
  In the case of M31, the best that we can do is to
  use the slope derived by the larger sample of \citet{sanders12},
  thus with a better statistics, rescaling the value of the intercept
  by the average difference of the two bins in common between
  \citet{ZB12} and \citet{sanders12}.  
In the bottom panel  we compare the PN and
H{\sc ii} region gradients. As in the case of NGC300, and even more
evidently, the gradient  of H{\sc ii} regions is steeper that of the one of the PN
population. However, we recall that for this galaxy we are able to 
estimate the effects of radial migration. These effects are 
particularly important at $\sim 8$~kpc, corresponding to $\sim 0.4 R/R_{25}$  in the $R_{25}$ scale.
This is exactly where we observe an increase in the 
scatter of the PN abundances and where the radial gradient of PNe 
show a ``jump'' of about $\sim 0.2$~dex. Thus, the steepening 
with time of the slope of the gradient and the large scatter could be, in part, attributed 
to the radial migration of its PN population. 
  
{\em M81.} The results  for M81 are  shown in Fig.~\ref{M81}. 
As for the other galaxies examined, the PN gradient is flatter than that of the H{\sc ii} regions, as already noticed by \citet{S14}. 
As for NGC300, no estimate of radial migration for this galaxy are available, thus the steepening with time can be attributed both 
to migration and to chemical evolution.  

\subsection{General findings}

We summarise  all gradients  in Table~5.
Our four galaxies have gradient slopes from the H{\sc ii} region populations ranging from $-0.22\, {\rm dex}/(R/R_{25}$)
in M33 to $-0.75\, {\rm dex}/(R/R_{25}$) in M81; PN gradients are generally flatter, from $-0.08\, {\rm dex}/(R/R_{25}$) in M31 to 
$-0.46\, {\rm dex}/(R/R_{25}$) in M81.

\begin{figure}
\centering
\includegraphics[width=0.35\textwidth, angle=270]{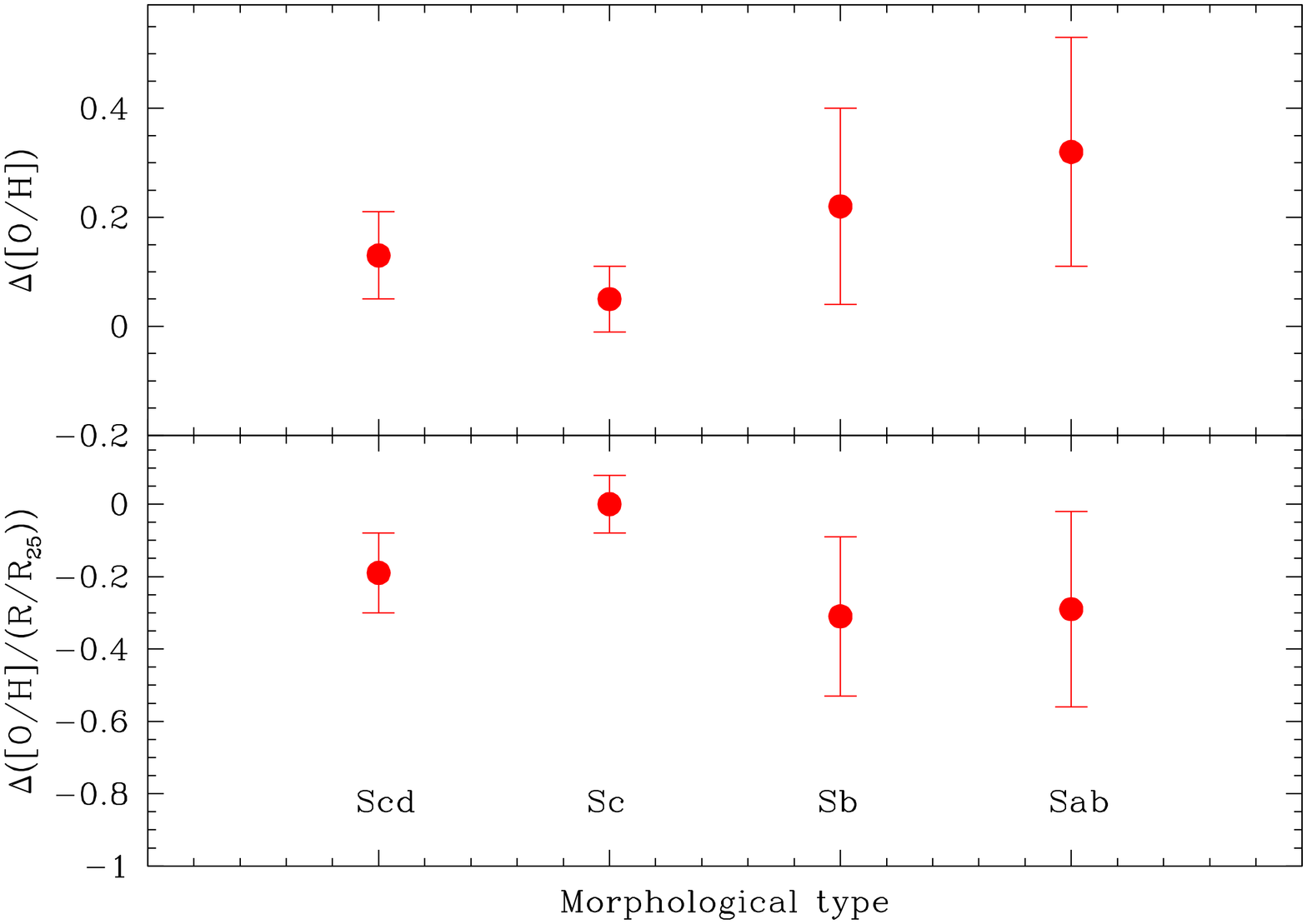}
\caption{Global enrichment (upper panel) and variation of the slope
of the metallicity gradient (lower panel) as a function of the
morphological type.} 
\label{enrichment} 
\end{figure}

The first consideration regards the global enrichment of the
galaxies, which can be approximated by the difference in the
value of the intercepts in Table~\ref{tab_meanabu}. The four galaxies
have all increased their metallicity content  from the epoch of the
formation of PN progenitors to the present time. For M31 the intercept
computed using the strong-line abundances of \citet{sanders12}
cannot be directly compared to the PN abundances. Thus we have
corrected it using the abundances obtained with the direct method
by \citet{ZB12}: the new value gives an almost solar abundance.
The metallicity enrichment  range from $+0.05$~dex in M33  to $+0.32$~dex in M81.  
The results of
the global enrichment are presented in the upper panel of
Fig.~\ref{enrichment}, where the variations of the intercepts versus
the morphological type are shown.  Late morphological types have
a smaller enrichment while a higher enrichment occurs
in early morphological types.


All galaxies of our sample present a small evolution of the
gradient slope, that can be a simple uniform metallicity increase
as in M33, or a steepening, as in NGC300 and  M31. From the present data,  there is no
observational evidence of gradients flattening with time.  In the
lower panel of Fig.~\ref{enrichment} we present the variation of
the gradient slope versus the morphological type.  The most
significant evolution occurs in early morphological types, while
late types tend to have no or  smoother evolution.
However, we remind that in the case of M31 the radial migration of PNe have an important role in shaping  the 
past gradient. 

\begin{table}
\caption{Gradients of the binned H{\sc ii} regions and PN populations.}
\begin{tabular}{llll}
\hline
Galaxy & Population & Slope  & [O/H] Intercept\\
 & & dex/($R/R_{25})$  &  \\
\hline
NGC300 & H{\sc ii} & $-0.29\pm 0.09$ & $-0.19 \pm 0.04$\\
	       & PN & $-0.10 \pm 0.07$ &  $-0.32 \pm 0.04$ \\
\hline
M33  &H{\sc ii} & $-0.22 \pm 0.06$ &  $-0.27 \pm 0.03$ \\
	 &PN & $-0.22 \pm 0.05$  &  $-0.32 \pm 0.03$ \\
\hline
M81 	       & H{\sc ii} & $-0.75 \pm 0.10$ & $+0.18 \pm 0.04$ \\ 
                &PN & $-0.46\pm 0.25$ & $-0.14 \pm 0.17$ \\
\hline
M31 	       & H{\sc ii} & $-0.39\pm 0.14$ & $+0.47\pm 0.09$ \\
               & H{\sc ii}($^*$)  &  & $-0.01$ \\
	       & PN & $-0.08 \pm 0.17$ &  $-0.22 \pm 0.11$\\
\hline
\end{tabular}
\\($^*$) corrected for {\em direct} method.
\label{tab_meanabu}
\end{table}

\section{Discussion and comparison with models}
\label{sec6} 

 It is important to compare our results with different kind of galactic chemical evolution models. 
We have selected two type of models: the classical chemical evolution models, specifically those of the grid of \citet{molla05}, and 
the cosmological models
of \citet{gibson13} with different treatment of the feedback.

\begin{figure}
\centering
\includegraphics[width=0.38\textwidth, angle=270]{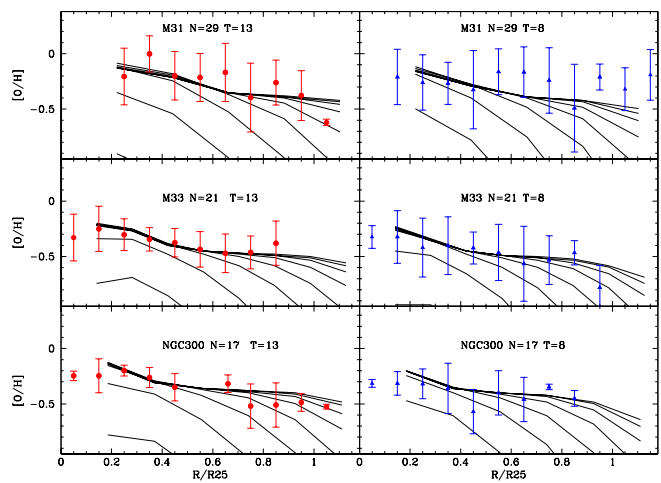} 
\caption{Comparison with the multi-phase chemical evolution models
of \citet{molla05} for galaxies with different mass and star formation
efficiencies.  The  panels to the left  present the models at the
present time ($T=13$~Gyr), while the panels to the right show the
models at the epoch of PN formation ($T=8$~Gyr).  Each curve is for a 
given star formation efficiency. For each galaxy we chose the most appropriate mass
distribution indicated by N in the plots,   where N is the designation of the model in Table~1 of \citet{molla05}}.
\label{All_evo}
\end{figure}

In Fig.~\ref{All_evo} we compare the gradients of the 
galaxies with the multi-phase chemical evolution models of \citet{molla05}
for galaxies with different mass and star formation efficiencies. We excluded M81 because of the strong interaction and environmental effects that 
prevent to compare with models of isolated galaxies. 
We selected the most appropriate model  on the basis of the total mass of each galaxy,  including dark matter,  from
the grid of  \citet{molla05}, varying the star formation
efficiency.  The adopted total masses are: 2.2$\times$10$^{11}$~M$_{\odot}$, 4.3$\times$10$^{11}$~M$_{\odot}$, and 2.0$\times$10$^{12}$~M$_{\odot}$ for NGC300, M33, and M31, respectively 
\citep{molla05}. 

  In the figure we indicate with N the model of  Table~1 of \citet{molla05} to which
we are referring, while the different curves corresponds to different sets of molecular cloud and star formation efficiencies (ranging from  0.007 to 0.95 for the cloud formation efficiency and to 4$\times$10$^{-6}$ to 0.088 for the star formation efficiency). 

Left-hand panels show the present-time models and
H{\sc ii} region abundances, while right-hand panels show models at roughly the
time of the formation of the oldest PN progenitors,  i.e., about 5 Gyr ago, and PN abundances.
In general, the agreement between the H{\sc ii} region gradients and the
present-time modelled gradients is satisfactory for the models with
the highest star formation efficiency  (from 0.14 to 0.88).  Lower star formation
efficiencies  (from 0.05 to 4$\times$10$^{-6}$) produce gradients  with shorter scale-lengths.
For NGC300 and M33, the observed PN gradients and the
models are quite close, whereas for M31 the result from the old
population is flatter than any modelled galaxy.  In this galaxy, radial migration 
is important and can contribute to the flattening of the PN gradient. 

An interesting point
to be noticed from Fig.~\ref{All_evo}  is that modelled radial
metallicity gradient, and likely the {\em real} gradients, are well
away from being approximated by a single slope.  Probably, our
simplistic assumption of a single slope gradient hides a more
complex time evolution of the metallicity distribution, in which,
for instance, inner and outer regions do not evolve at the same
rate.

In Fig.~\ref{All_evo_cosmo}, we compare the cosmological models
of \citet{gibson13} with different treatment of the feedback,
''enhanced''  and  ''conservative'' feedbacks, with our gradient
slopes from H{\sc ii} regions and PNe.  The models are built to reproduce
the present-time gradient of a Milky-Way-like galaxy, therefore it
is not surprising that the agreement with the H{\sc ii} regions gradient
is good.  The range of values for our H{\sc ii} region gradients
indicates the dependence on the morphological type when the gradients
are expressed in terms of dex~kpc$^{-1}$.  On the other hand, the gradients
traced by PNe put important constraints on the models: the current
data certainly favour a smooth evolution of the gradient and are
not consistent with a flatter gradient in the past in the time lapse
(and redshift interval) constrained by PNe.  Comparing the data
with the two sets of models, ''enhanced''  and  ''conservative''
feedbacks, at the epoch of birth of PNe we can discern a better
agreement with the first class of models, in which the energy of
SNe is widely redistributed.
All PN gradients in this plot have
been placed at 5~Gyr, corresponding to progenitor mass at turnoff
of 1.2~$M_{\odot}$  \citep{maraston98}. This is the lower limit of masses
of Type~II PNe, favoured by the IMF, but may not strictly represent
the progenitor mass/look-back time for all PN within each sample.
Each PN sample can be contamined by a few Type III PNe, which would
probe higher redshifts ($0.5<z<8$). On the other hand, we have excluded Type~I PNe,  as discussed above. 
With cosmological
parameters  $H_0=67.04$~km~s$^{-1}$~Mpc$^{-1}$, 
$\Omega_{\rm m}=0.3183$, $\Omega_{\Lambda}=0.6817$,
this look-back time corresponds to a redshift of $z \simeq 0.5$.

It is clear from the plot that  extragalactic PNe cannot trace
the most remote past of galaxies, probing the epochs where the two scenarios mostly differ. 
For example, one of the models with more conservative feedback (the dashed magenta  curve in Fig.~\ref{All_evo_cosmo})
is still consistent, within the errors, with the observational
constraints from PNe.  It is worth noting that Type~III Galactic
PNe probe look-back time up to 8~Gyr, and that the Type III PN
gradient observed in the Galaxy \citet{stanghellini10} also
agrees with the enhanced feedback models \citep[see][]{gibson13}.
Nonetheless, the Galactic PN gradient suffers from uncertainties
in the distance scale, which make this constraint not very strong.
For these reasons, Local Universe constraints need to be complemented
with high-redshift observations, as those of \citet{cresci10},
\citet{jones10, jones13, jones15}, and \citet{yuan11}, that are able 
to shed light to the very early phases of disc evolution.

\section{Summary} 
\label{sec7}
In the present work, we have analysed in a homogeneous way the
literature data on {\em direct}-method abundances of H{\sc ii} regions and PNe in four
nearby disc galaxies: NGC300, M33, M31, and M81.  
The abundances of H{\sc ii} regions in M31 are 
an important exception, since only very few regions have {\em direct}-method abundances \citep{ZB12}, while a large sample with abundances derived with strong lines
is available \citep{sanders12}. 
In this case,  we take advantage of the large sample of \citet{sanders12} and we use the sample \citep{ZB12}  to
compute the offset between the {\em direct} method and the strong-line
calibration.

The analysed galaxies belong
to different morphological types and allow us to study the time
evolution of the metallicity enrichment and of the gradient 
as a function of galaxy morphology  assuming that no correction is needed for the slope for stellar migration.  To properly exploit PNe as
tracers of the past composition of the ISM, we have proved that their
O abundances are unchanged at first order during the progenitor stellar
evolution.  Moreover, we have estimated the amplitude of radial
migration of the two galaxies in the sample with measured radial
velocities. In the case of M33, we have shown that the effects of
radial migration on the metallicity gradients are negligible. 
In the case of M31, the effects of
radial migration in the PN velocity field is non-negligible, and
could have affected significantly the radial metallicity gradient  in the
last few Gyr.

Using H{\sc ii} regions and PNe as tracers of the time evolution of the
metallicity distribution in disc galaxies, we have found that: ({\em
i}\/) all galaxies are subject to a global oxygen enrichment with time,
higher for earlier type spirals (Sa-Sb); ({\em ii}\/) on average, the
O/H gradients of the older population are equal to or flatter
than those of H{\sc ii} regions.  While steeper O/H gradients for young populations 
have been noted before in spiral galaxies from {\em direct}
abundances of PNe and H{\sc ii} regions \citep{magrini09b,stasinska13,S14}, 
we confirm this trend when galaxies are studied
in a comparable radial domain and metallicity binning.
({\em iii}\/) Our kinematical study of PNe in M33 and in M31 finds that radial migration can contribute to the observed flattening, 
especially in the case of M~31. The undisturbed PN population of M33 allows us to better constrain the time evolution 
of the radial gradient.

Our radial gradients and their time-evolution are  compared with classical and cosmological chemical evolution models. 
They are found to be in good agreement with the multiphase chemical evolution models  
by \citet{molla05}. 
Using our results as constraints to the cosmological models of
galaxy evolution of \citet{gibson13}, we find that, in the redshift range
sampled by H{\sc ii}~regions and PNe,  models with enhanced 
SN feedback are favoured.

\begin{figure}
\centering
\includegraphics[width=0.35\textwidth, angle=270]{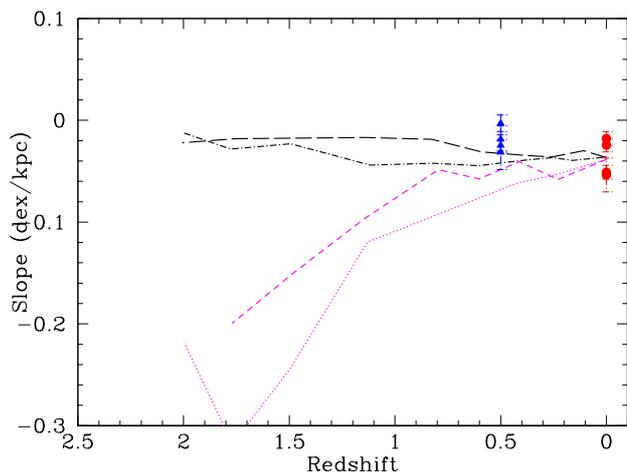} 
\caption{Comparison of the redshift evolution of the gradient
with the chemical evolution models of \citet{gibson13}: black (dashed and dot-dashed) lines
are models with enhanced feedback from SNe, magenta (short-dashed and dotted) lines are models with normal feedback. Red circles and blue 
triangles show the slopes of the gradients from H{\sc ii} regions 
and PNe with 1.2~$M_\odot$
progenitors, respectively.  Note that in this plot the slopes are expressed in
dex/kpc to be compared with the models of  \citet{gibson13}. }
\label{All_evo_cosmo}
\end{figure}

With the current technology, the realm of star-forming galaxies, where the time evolution of the metallicity
gradient evolution can be obtained from {\em direct}-method
abundances, is limited. By setting a limit from the PNLFs studied so far and reasonable observability in a few
8m-class telescope nights, we expect to enlarge our sample
by few more galaxies. This would allow to refine the evolutionary trends shown
in Fig.~\ref{All_evo_cosmo}. We also expect to acquire in the near future deeper
spectra of the current galaxies, to better characterise their PN and  H~{\sc ii} region populations, as for instance in M31, 
and possibly to reveal older PN populations. 
However, the strongest constraints 
to the galactic evolutionary models are expected to come from either 
a better determination of the distance scale for
Galactic PNe, to which the Gaia satellite will contribute, from abundance studies of high-redshift galaxies, and from 
the future 30m-class telescopes. 
The latter will favour both better characterisation of PNe and H{\sc ii} regions in the nearby galaxies
with optical multi-object spectroscopy and  abundances of redshifted galaxies with infrared integral field unit spectroscopy.

\begin{acknowledgements}
The authors are very grateful to the referee for her/his constructive report that improved the quality and presentation of the paper. 
This research has made use of the NASA/IPAC Extragalactic Database (NED) which is operated by the Jet Propulsion Laboratory, 
California Institute of Technology, under contract with the National Aeronautics and Space Administration. 
We acknowledge the usage of the HyperLeda database.
V.C. acknowledges DustPedia, a collaborative focused research project supported by the European Union under the Seventh Framework Programme (2007-2013) call (proposal no. 606824). The participating institutions are: Cardiff University, UK; National Observatory of Athens, Greece; Ghent University, Belgium; Universit\'e Paris Sud, France; National Institute for Astrophysics, Italy and CEA (Paris), France. 
\end{acknowledgements}

\Online

\begin{appendix} 
\section{Circular model fit of M31}
In this appendix we show the circular model fit for all the elliptical
bins in M31 as performed in Section \ref{sec:migrating2}.

\begin{figure*}
\centering
\vbox{
  \hbox{
    \includegraphics[width=0.3\textwidth, angle=0]{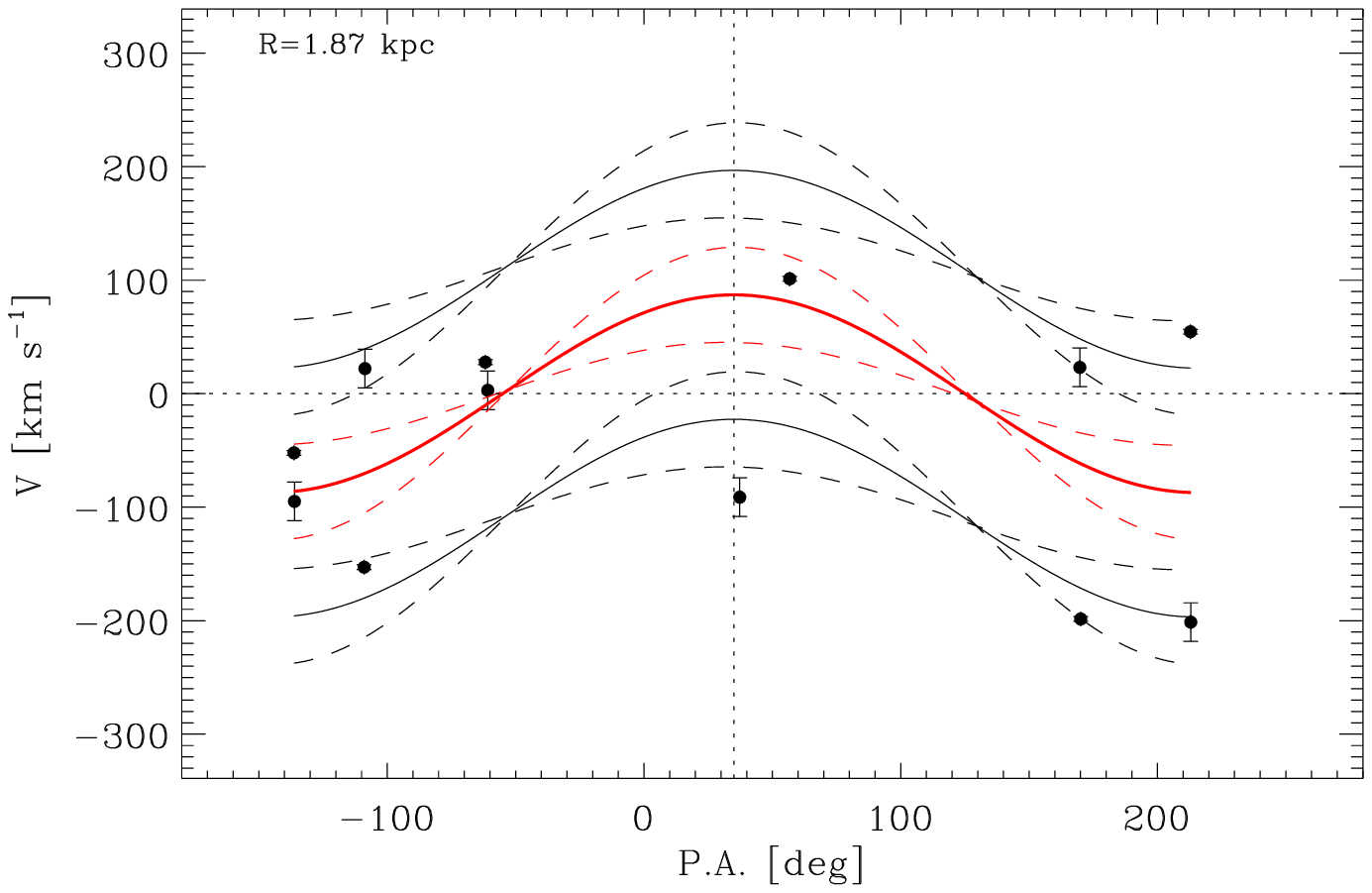}
    \includegraphics[width=0.3\textwidth, angle=0]{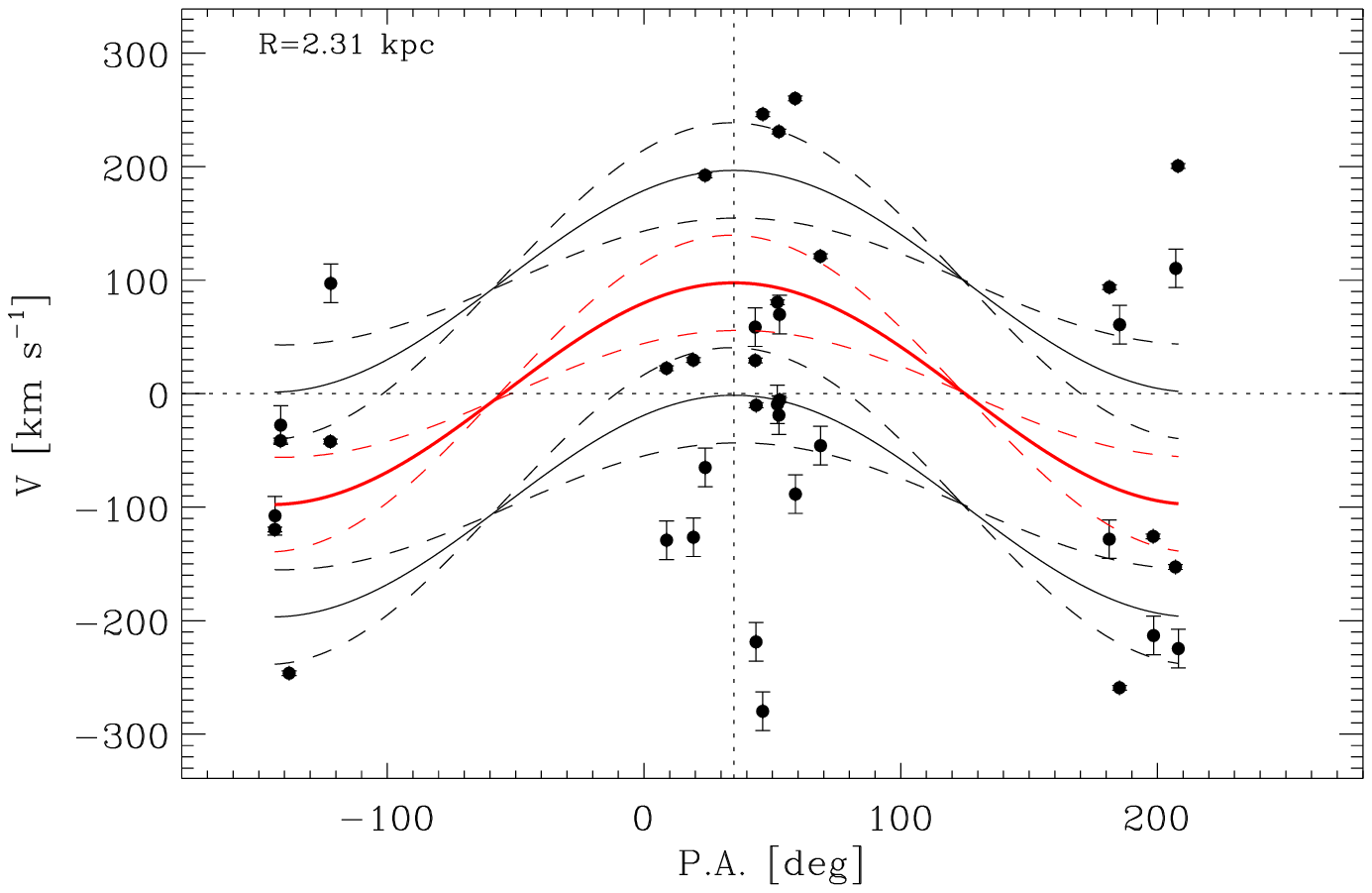}
    \includegraphics[width=0.3\textwidth, angle=0]{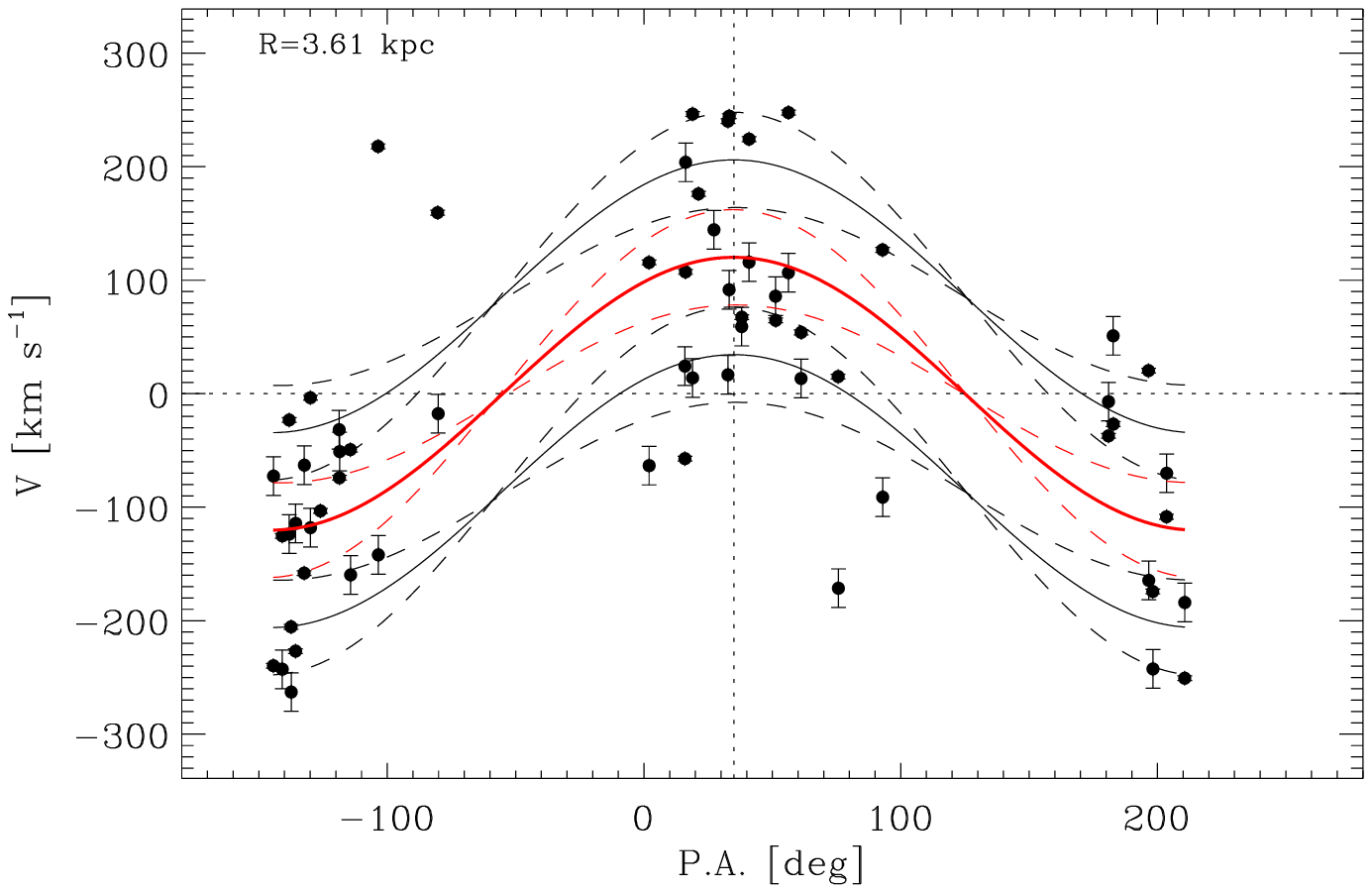}
  }
  \hbox{
    \includegraphics[width=0.3\textwidth, angle=0]{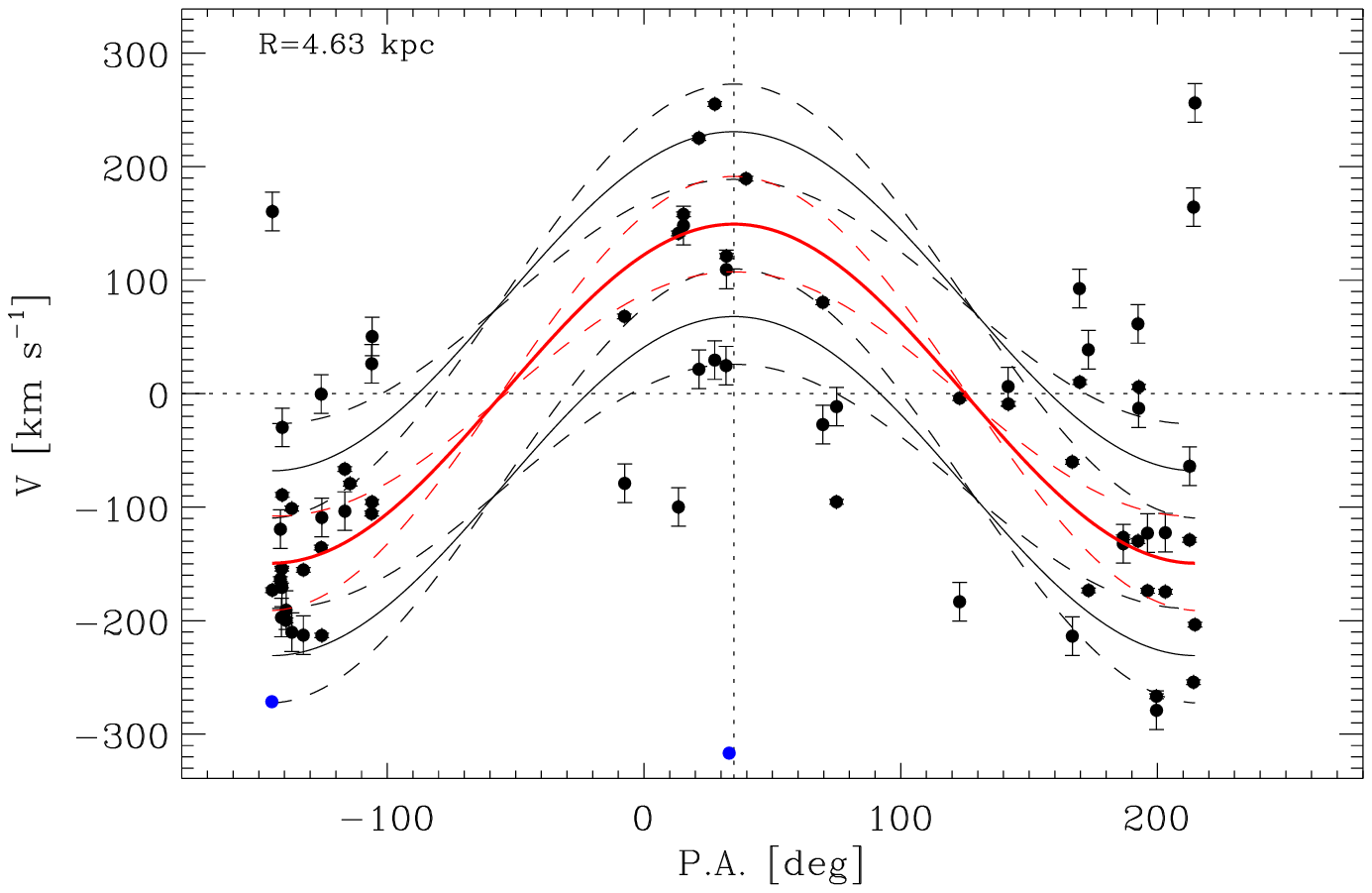}
    \includegraphics[width=0.3\textwidth, angle=0]{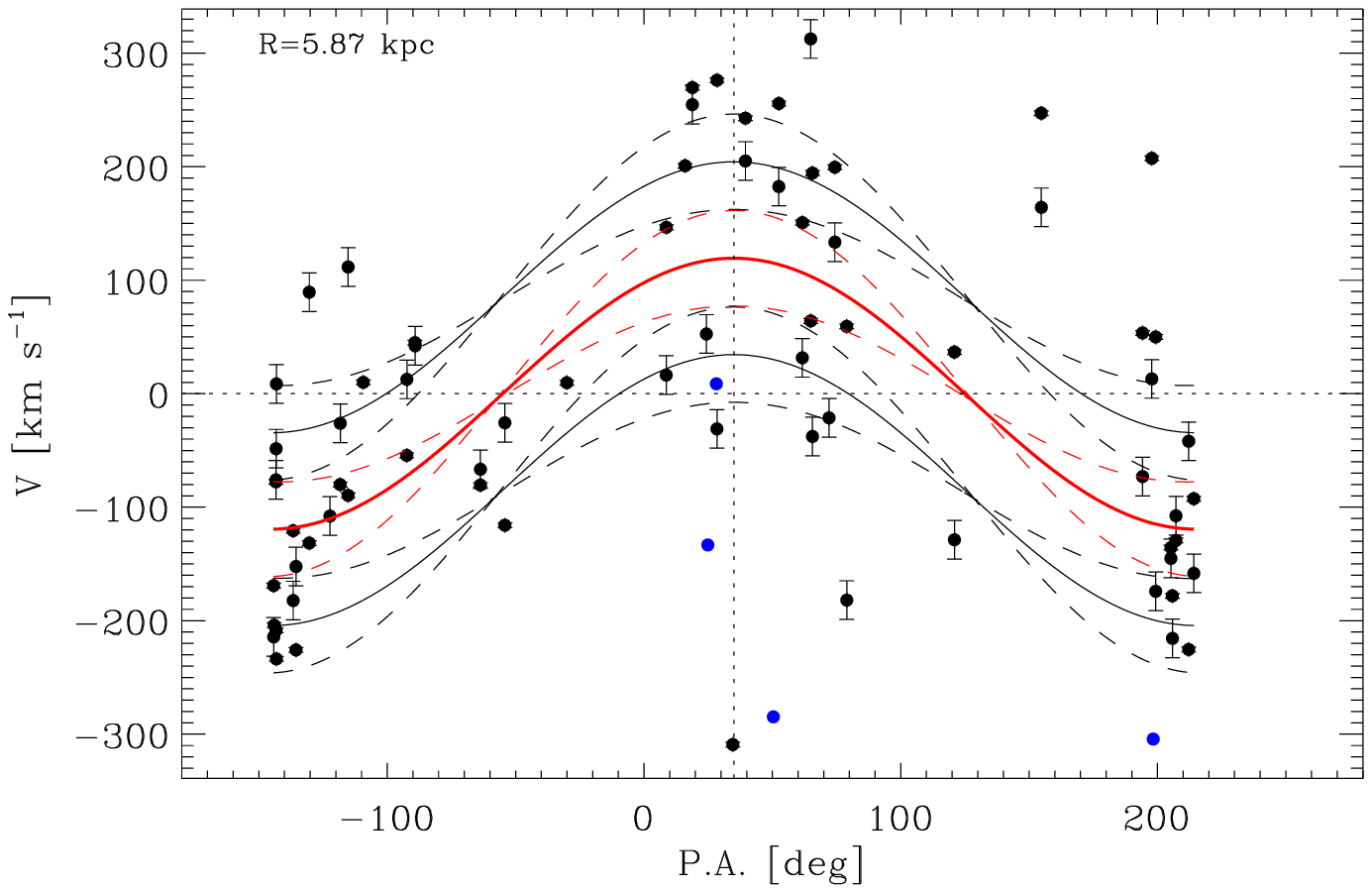}
    \includegraphics[width=0.3\textwidth, angle=0]{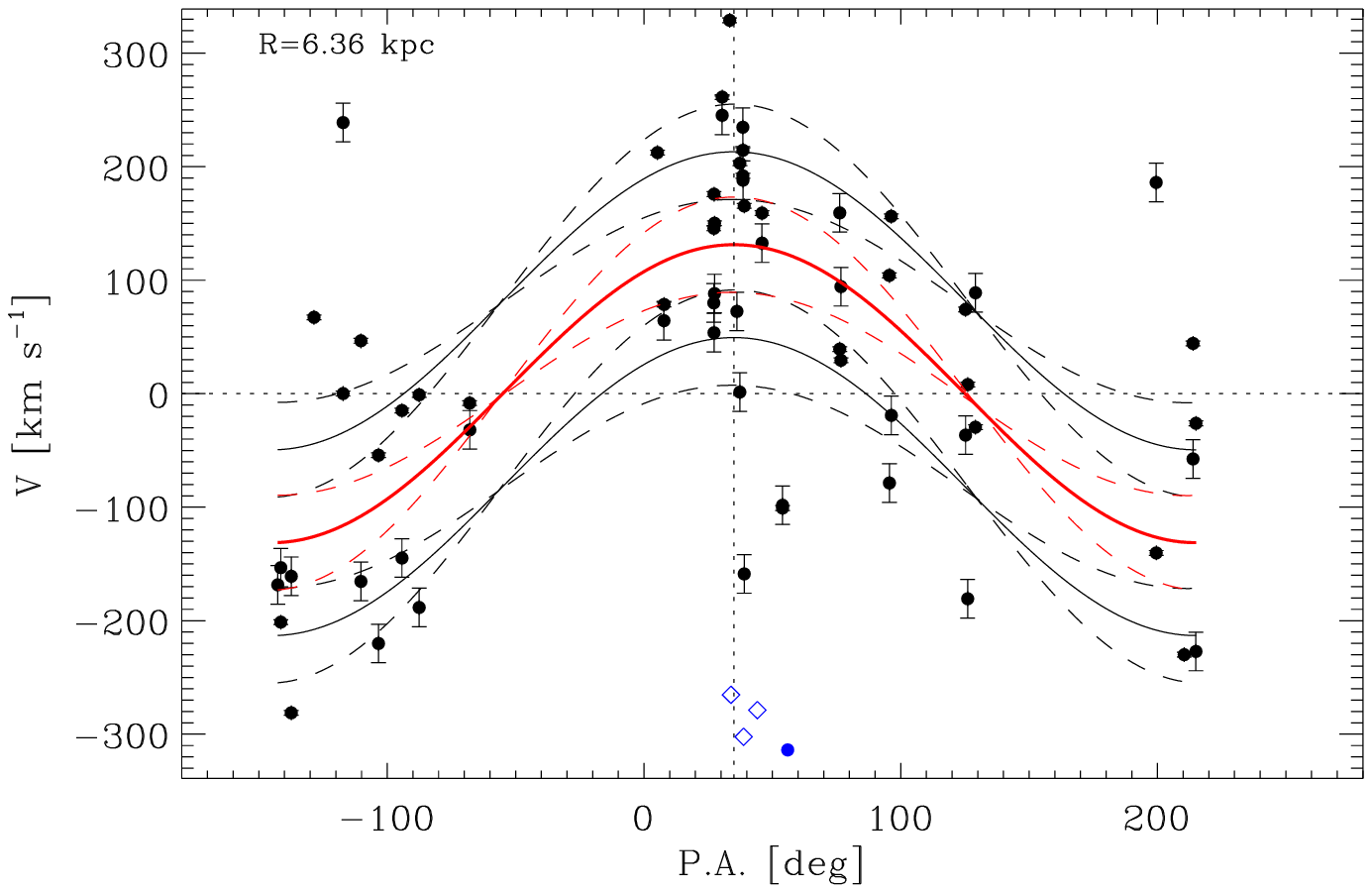}
  }
  \hbox{
    \includegraphics[width=0.3\textwidth, angle=0]{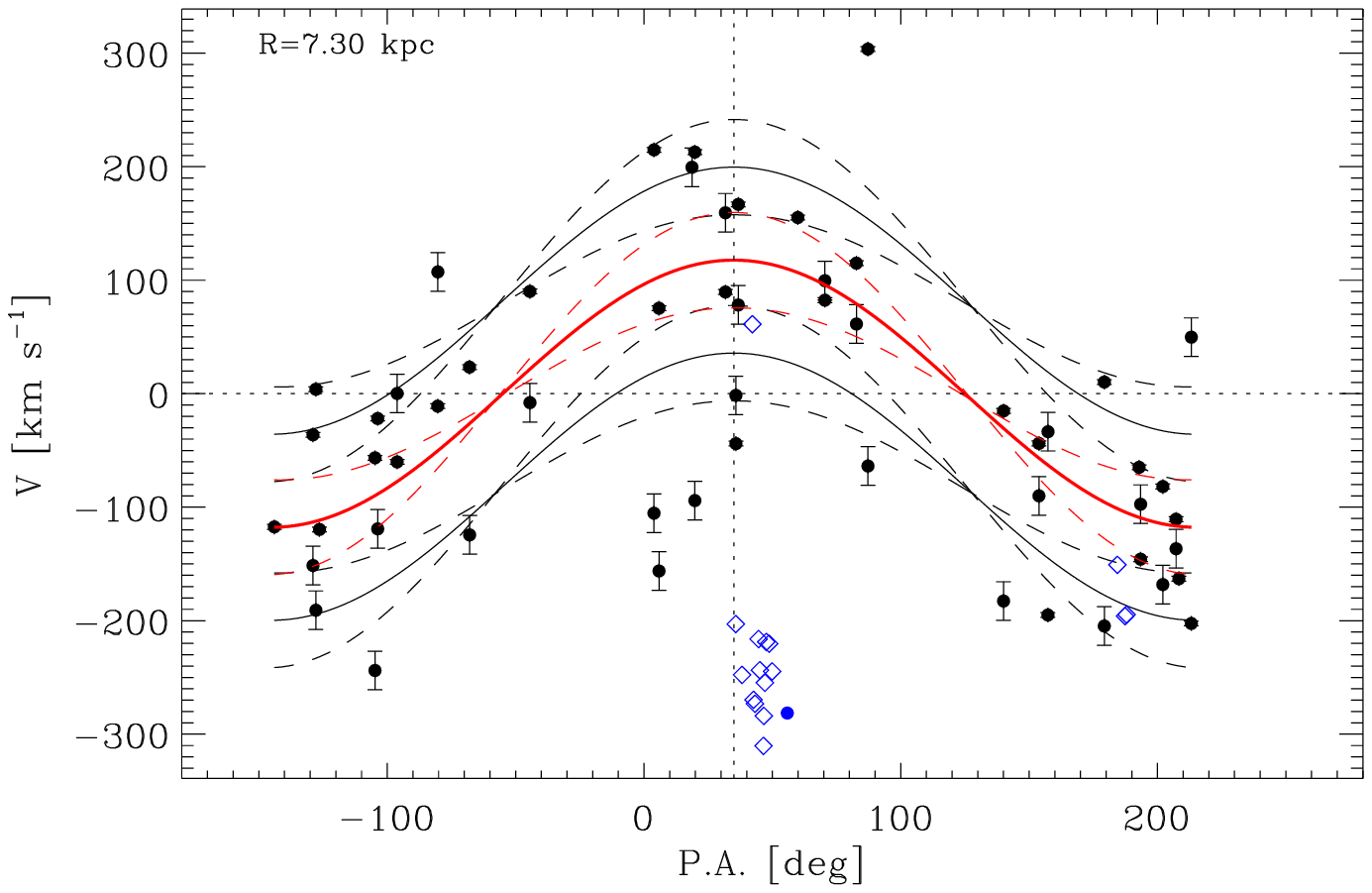}
    \includegraphics[width=0.3\textwidth, angle=0]{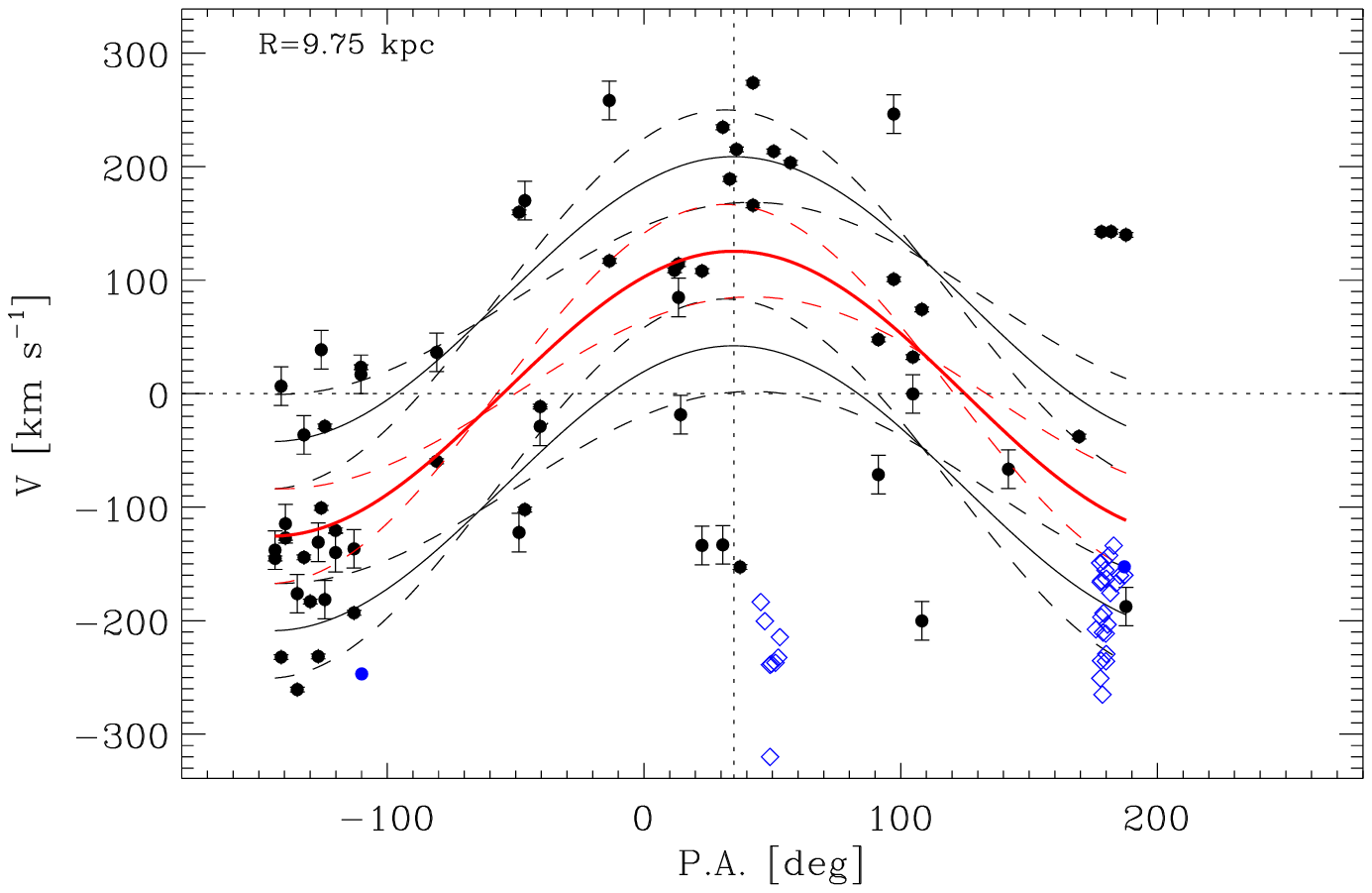}
    \includegraphics[width=0.3\textwidth, angle=0]{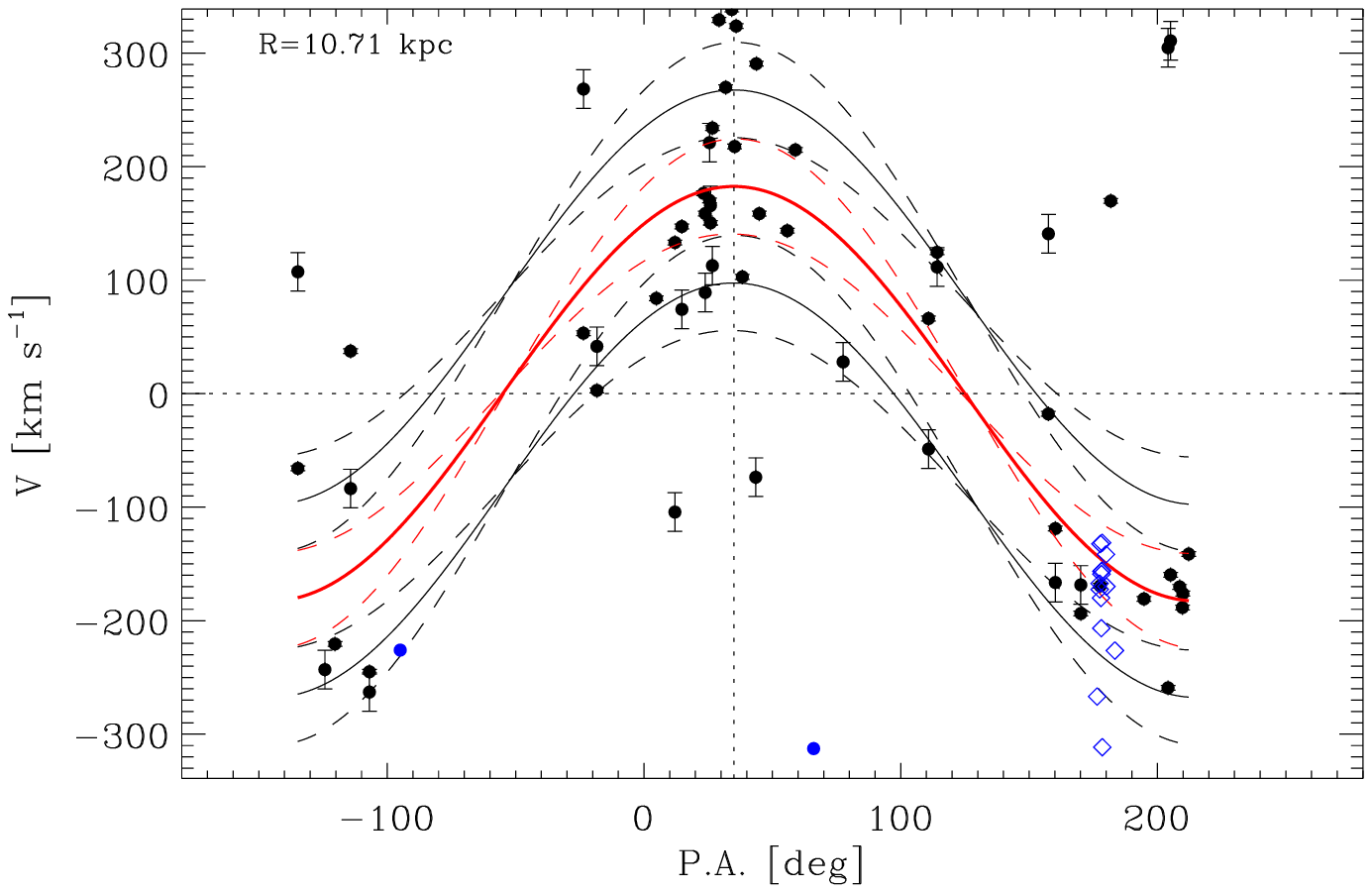}
  }
  \hbox{
    \includegraphics[width=0.3\textwidth, angle=0]{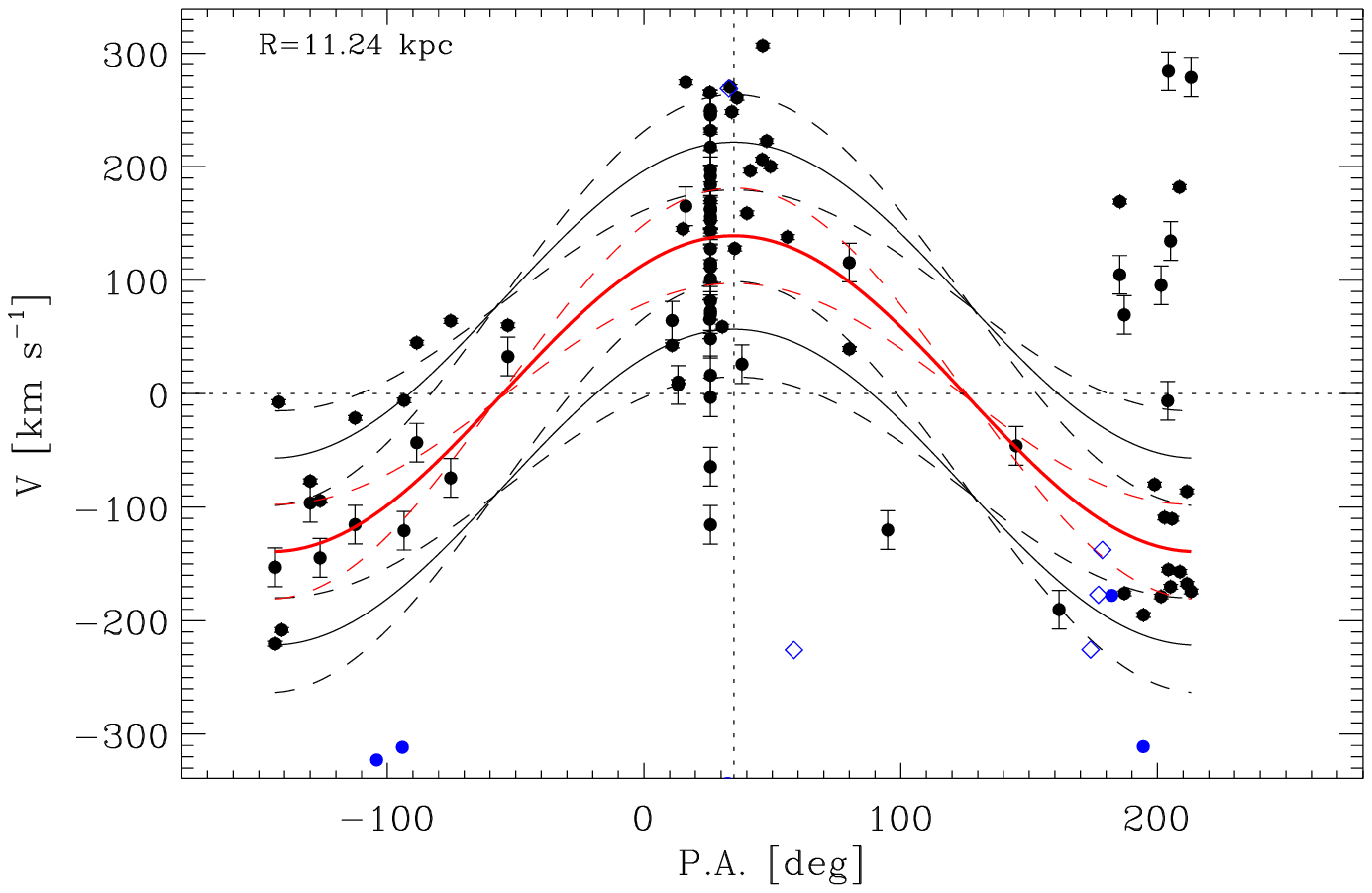}
    \includegraphics[width=0.3\textwidth, angle=0]{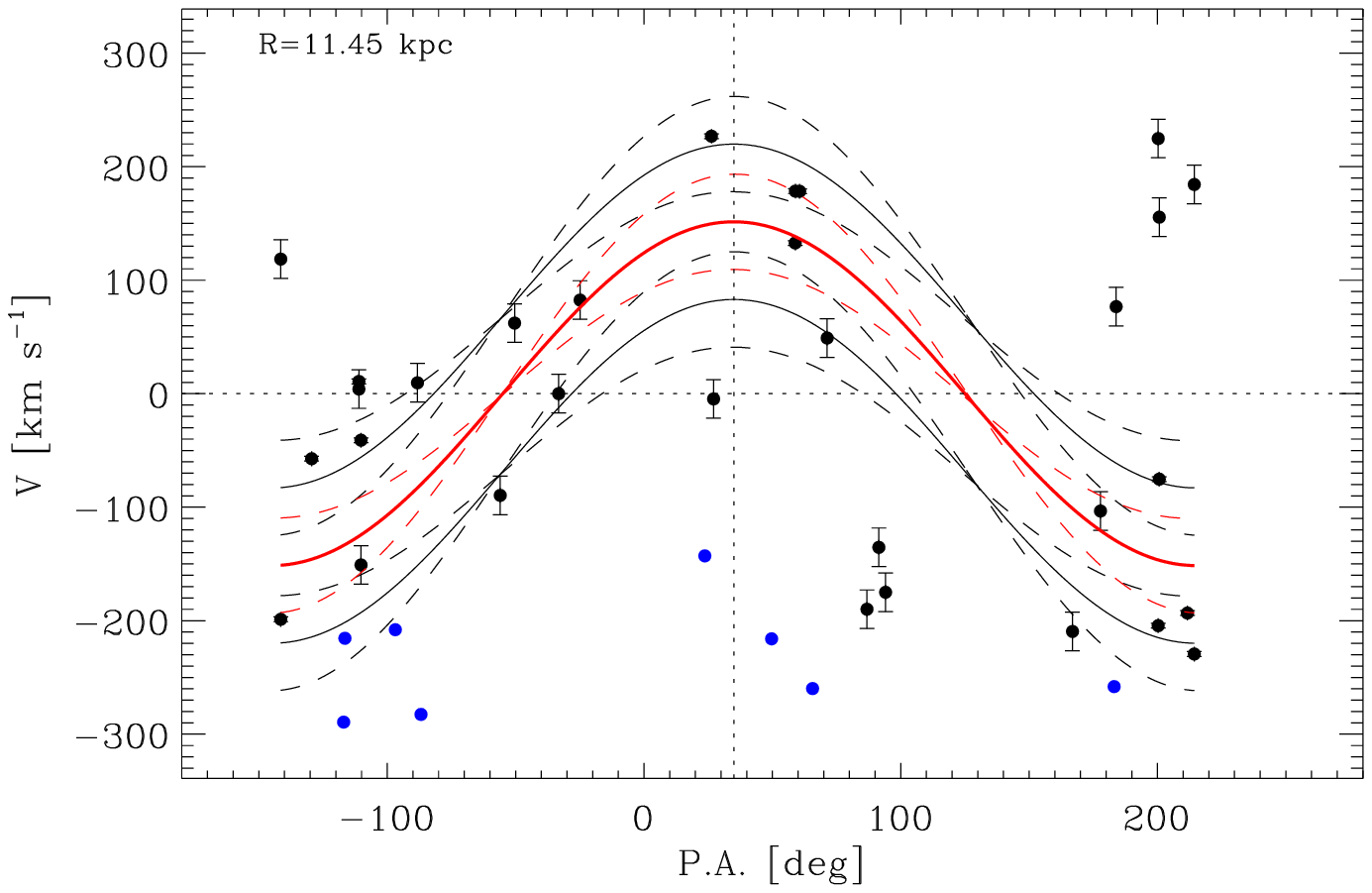}
    \includegraphics[width=0.3\textwidth, angle=0]{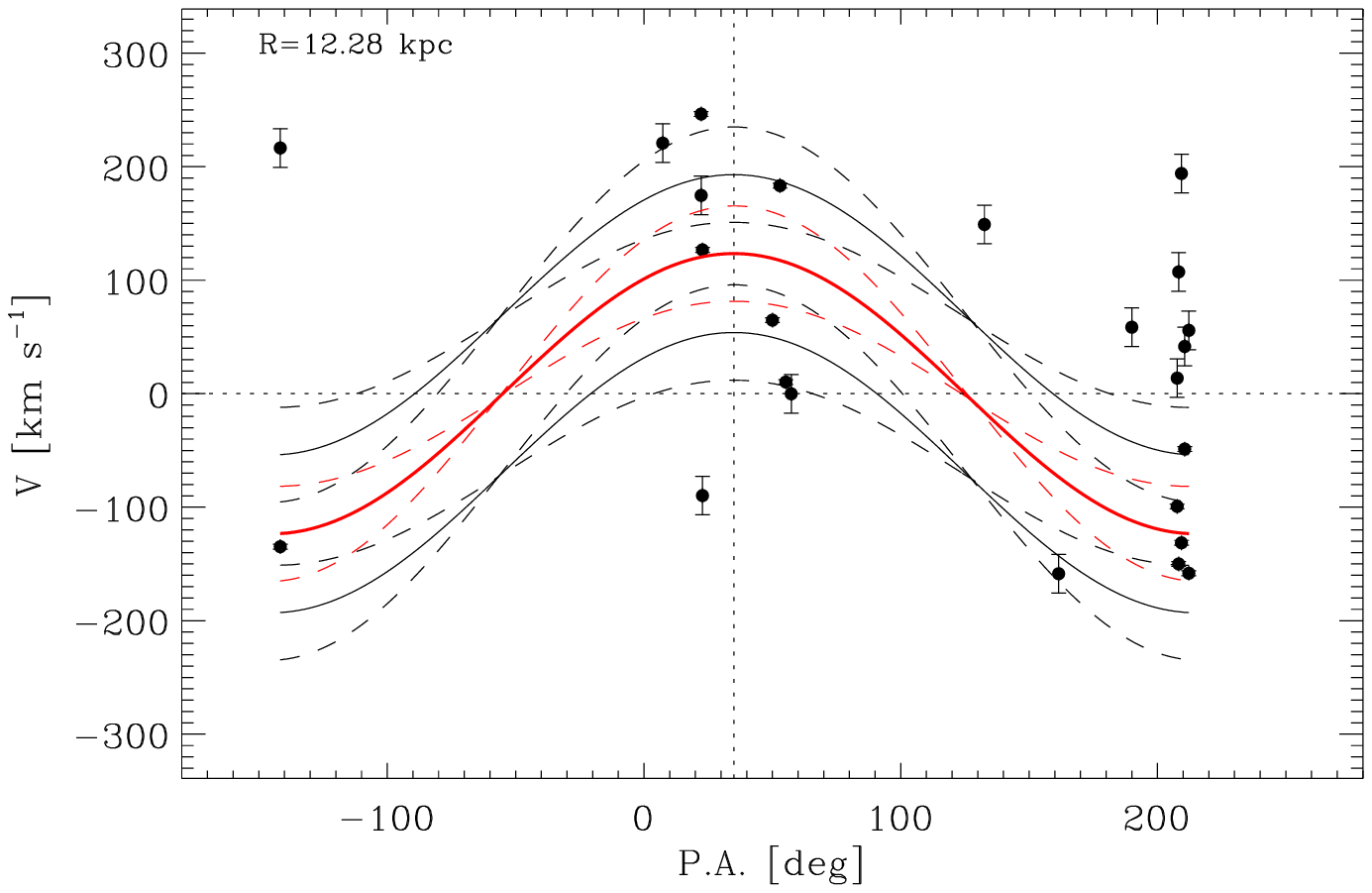}
  }
  \hbox{
    \includegraphics[width=0.3\textwidth, angle=0]{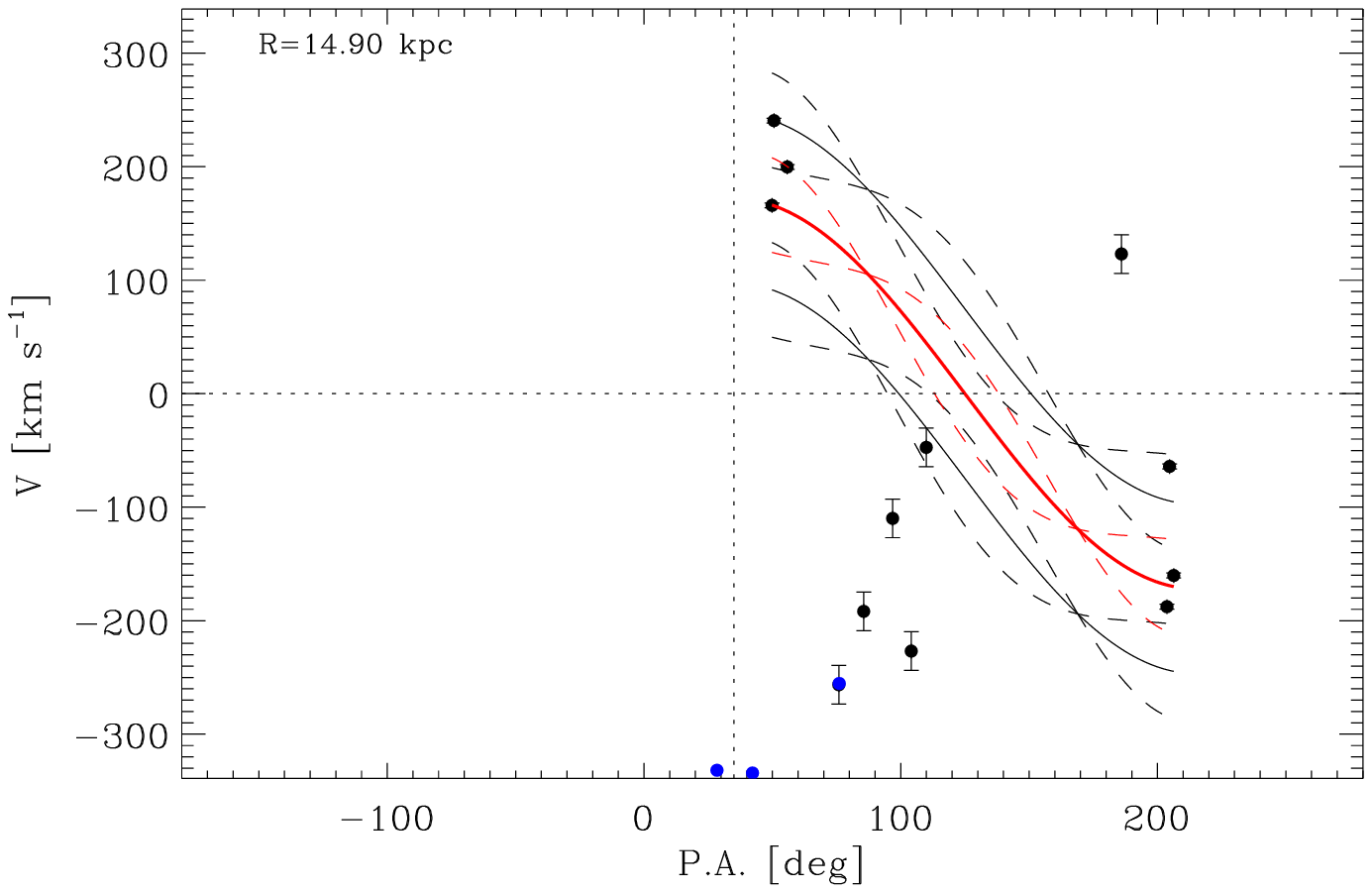}
    \includegraphics[width=0.3\textwidth, angle=0]{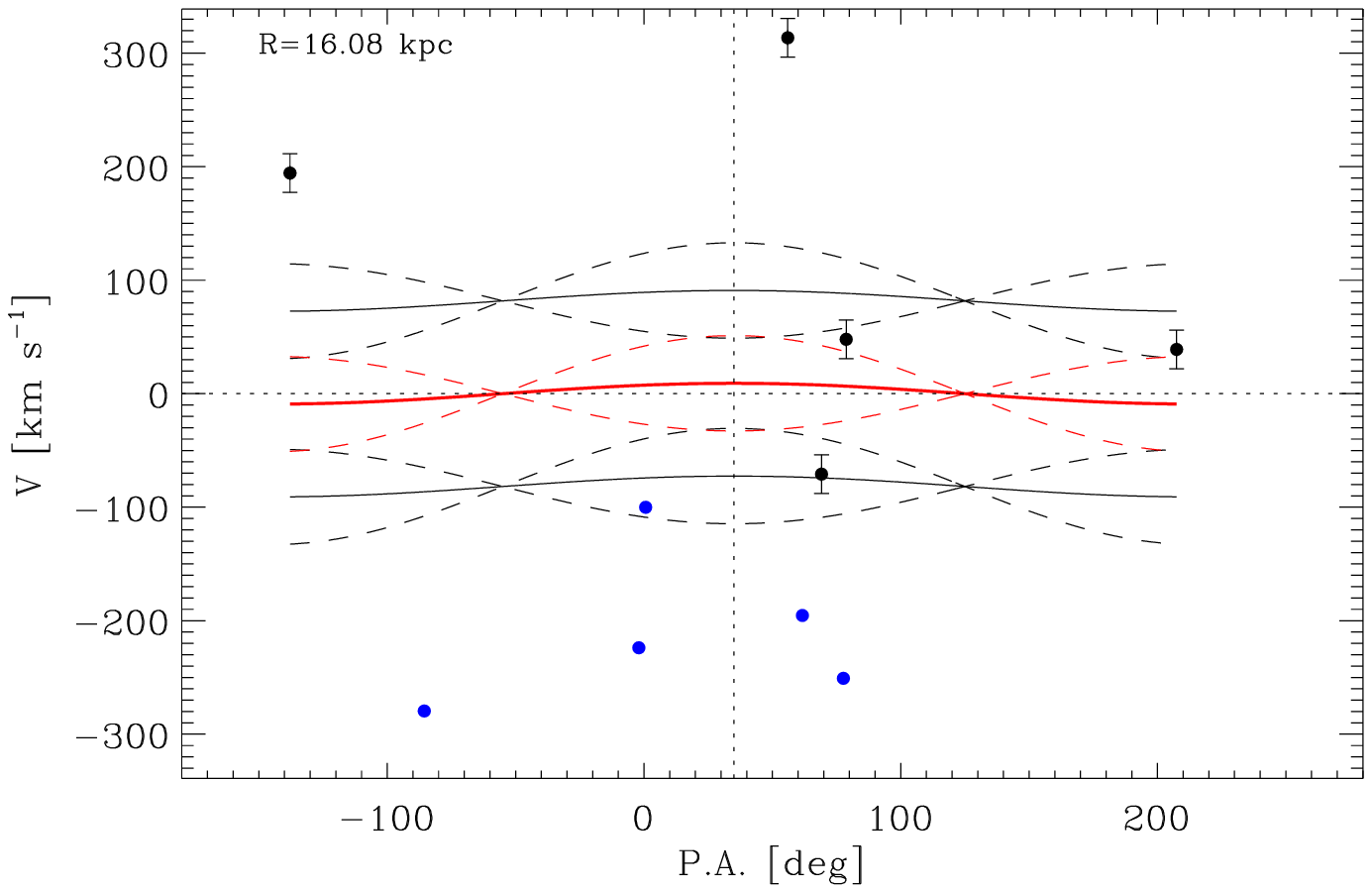}
  }
}
\caption{Same as Fig.\ref{fig:fit_bin31}, but for
  the remaining elliptical bins of M31.}
\label{fig:appendixA}
\end{figure*}

\end{appendix}
\end{document}